\documentclass[floatfix, superscriptaddress, twocolumn, showpacs, aps, pra, 10pt,amsmath,amssymb]{revtex4-2}
\usepackage{graphicx}
\usepackage{dcolumn}
\usepackage{bm}
\usepackage{stackrel}
\usepackage{amsmath}
\usepackage{amsfonts}
\usepackage{amssymb}
\usepackage{amsthm}
\usepackage{epstopdf}
\usepackage{longtable}
\usepackage{array}
\usepackage{pifont}
\usepackage{latexsym}
\usepackage[latin1]{inputenc}
\usepackage{textcomp}
\usepackage{subequation}
\usepackage{subfigure}
\usepackage{braket}
\usepackage{wasysym}
\usepackage{ulem}
\usepackage{color, soul}
\usepackage{multirow}
\usepackage{ragged2e}
\usepackage{fancybox}
\usepackage{mathtools}
\usepackage{rotating}
\usepackage{tikz}
\usepackage{hyperref}
\usetikzlibrary{shapes}
 \pdfoutput=1
\newcolumntype{.}{D{.}{.}{-1}}
\tikzset{hexagon/.style={regular polygon,regular polygon sides=6},}

\begin{document}
\title{Fault-tolerant Hyper-Ramsey spectroscopy of optical clock transitions\\ with dynamical decoupling}
\author{B. Ilikj*}
\thanks{These authors contributed equally to this work.}
\email{\\biilich@uni-sofia.bg\\thomas.zanon@sorbonne-universite.fr}
\affiliation{Centre for Quantum Technologies, Department of Physics, Sofia University, 5 James Bourchier Boulevard, 1164 Sofia, Bulgaria}
\author{T. Zanon-Willette*}
\affiliation{Sorbonne Universit\'e CNRS, MONARIS, UMR 8233, F-75005 Paris, France}
\author{D. Wilkowski}
\affiliation{MajuLab, International Research Laboratory IRL 3654, Universit\'e C\^ote d'Azur, Sorbonne Universit\'e, National University of Singapore, Nanyang Technological University, Singapore,}
\affiliation{Center for Quantum Technologies, Nanyang Technological University Singapore, 50 Nanyang Avenue, Singapore, 639798, Singapore.}
\affiliation{Nanyang Quantum Hub, School of Physical and Mathematical Sciences, Nanyang Technological University, 21 Nanyang Link, Singapore, 637371, Singapore.}
\author{B. Darqui\'e}
\affiliation{Laboratoire de Physique des Lasers, CNRS, Universit\'e Sorbonne Paris Nord, Villetaneuse, France}
\author{N.V. Vitanov}
\affiliation{Centre for Quantum Technologies, Department of Physics, Sofia University, 5 James Bourchier Boulevard, 1164 Sofia, Bulgaria}

\begin{abstract}
A probe laser field that interrogates an atomic or molecular transition also shifts it. Although Hyper-Ramsey (HR) spectroscopy was developed to mitigate these light-induced frequency shifts, the technique remains sensitive to probe-intensity-fluctuations. We introduce a novel class of dynamically decoupled hyper-Ramsey (DDHR) sequences that utilize a modified refocusing pulse to effectively eliminate this residual sensitivity. A three-pulse interferometric protocol significantly enhances the contrast of interference fringes while providing robust immunity against probe-induced frequency shifts, even in the presence of external field inhomogeneities. We experimentally validate both HR and DDHR protocols on NQCH and IQM superconducting quantum processors, demonstrating a strong reduction of frequency shifts rigorously consistent with frequency-scaled simulations for optical clocks. DDHR spectroscopy yields superior suppression of residual probe-induced frequency shifts compared to the standard HR scheme under pulse technical defects and decoherence. Furthermore, we demonstrate that composite refocusing pulses, derived from a Pascal binomial tree architecture, gives substantial resilience against phase-step imperfections during application of refocusing pulses. Ultimately, fault-tolerant, dynamically decoupled hyper-clocks will offer a promising experimental platform for the evaluation of more advanced sequences of periodic, non-periodic and concatenated multiple refocusing pulses holding significant potential for high-precision matter/antimatter quantum sensing and rigorous tests of fundamental physics within harsh electromagnetic environments.
\end{abstract}

\maketitle

\indent By engineering the interaction between light and matter, we can create robust optical qubits that resist environmental noise and imperfections. This approach is set to advance the fields of quantum simulation, computation, and metrology while developing fault-tolerant quantum sensors that open new avenues for exploring fundamental symmetries in physics, and the search for new physics beyond the Standard Model~\cite{Ye:2024,Bass:2024,Degen:2017,Safronova:2018}. Atomic optical clocks~\cite{Ye:2008,Derevianko:2011,Ludlow:2015} are a prime example of a highly active research field that is currently facing a new and challenging task in not only evaluating optical frequencies at a fractional inaccuracy at a relative level of $10^{-19}$~\cite{Brewer:2019,Aeppli:2024,Marshall:2025,Zhang:2026} but also in realizing highly stable clock lasers with record low phase and intensity noise~\cite{Dorscher:2020,Yan:2025,Zaporski:2025}. Spectroscopic performances can be enhanced through the optimization of control methods based on periodic driving, composite pulses and dynamical decoupling techniques used in quantum information processing ~\cite{Levitt:1986,Genov:2017,Viola:1998,Viola:2003,Viola:2004,Vandersypen:2005,Jones:2024,Tayler:2025}. These techniques provide efficient quantum engineering solutions to reduce shifts resulting from specific spin-spin interactions and distortions from many-body effects in experimental atomic physic platforms~\cite{Choi:2020,Zhou:2023}.
\begin{figure}[t!!]
\centering
\resizebox{8.6cm}{!}{\includegraphics[angle=0]{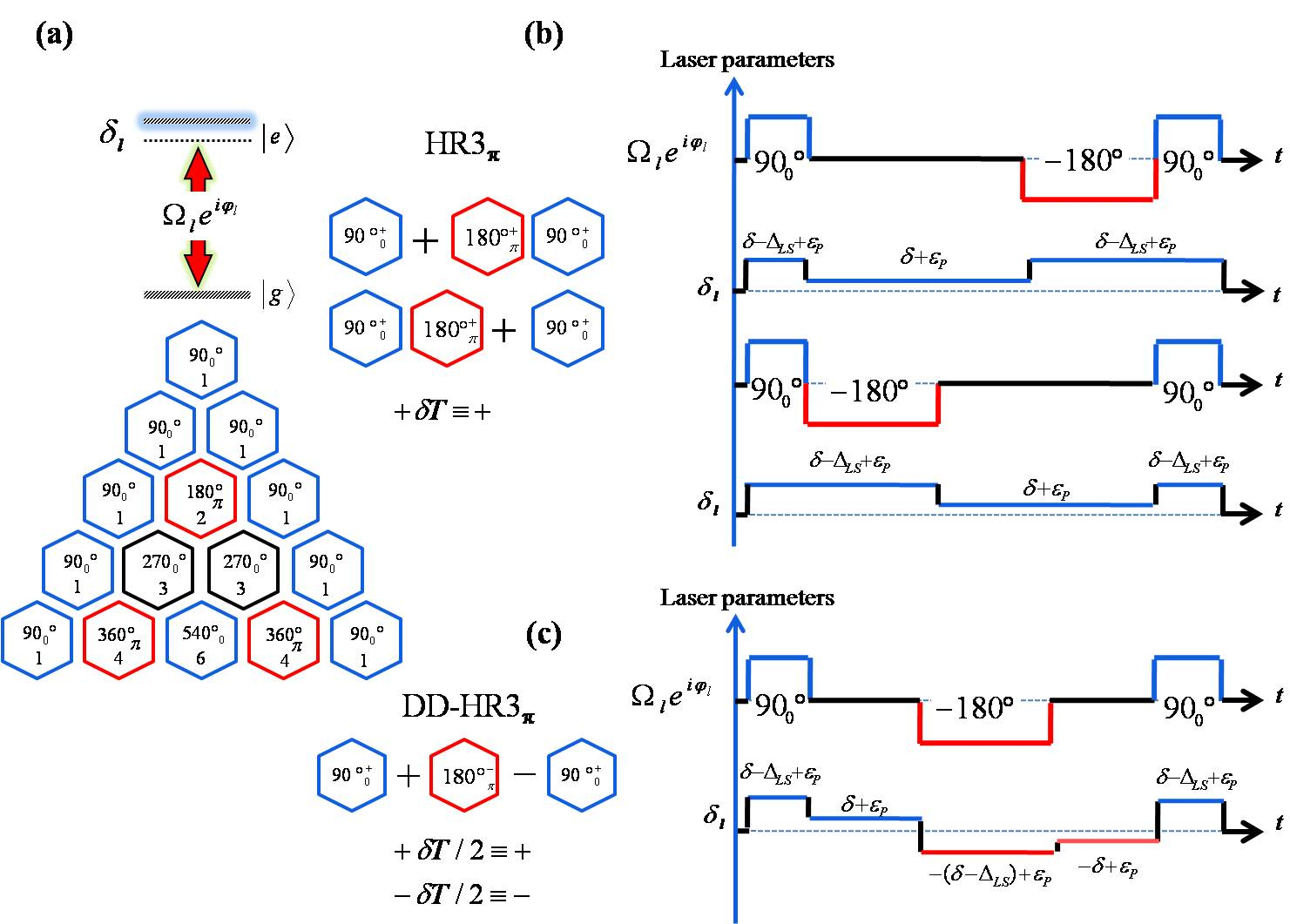}}
\caption{(a) Composite laser pulse protocols to probe a two-level quantum system (qubit) following a Pascal's triangle decomposition. (b)-(c) Classes of HR protocols (top) and DDHR protocols (down) based on phase-shifted refocusing pulses encapsulated by Ramsey pulses. Laser probe frequency detunings with opposite signs are indicated as $\pm$ exponents of the pulse areas and laser phase jumps as indices. Definitions of laser parameters are related to amplitude $\Omega_{l}$, phase $\varphi_{l}$ and frequency detunings of the l-th pulse number where $\delta_{l}=\pm(\delta-\Delta_{LS})$ (see schematics), $\delta$ is the frequency detuning between the laser probe and the qubit, with an additional light-shift correction $\Delta_{LS}$ and a potential probe frequency drift (or offset) $\varepsilon_{P}$. Free evolution zones without laser light are introduced by $+\delta T\equiv+$ for the HR protocol and $\pm\delta T/2\equiv\pm$ for the DDHR protocol.}
\label{fig:DD-HR-protocols}
\end{figure}
\begin{figure}[t!]
\centering
\includegraphics[width=8cm, height=5.5cm]{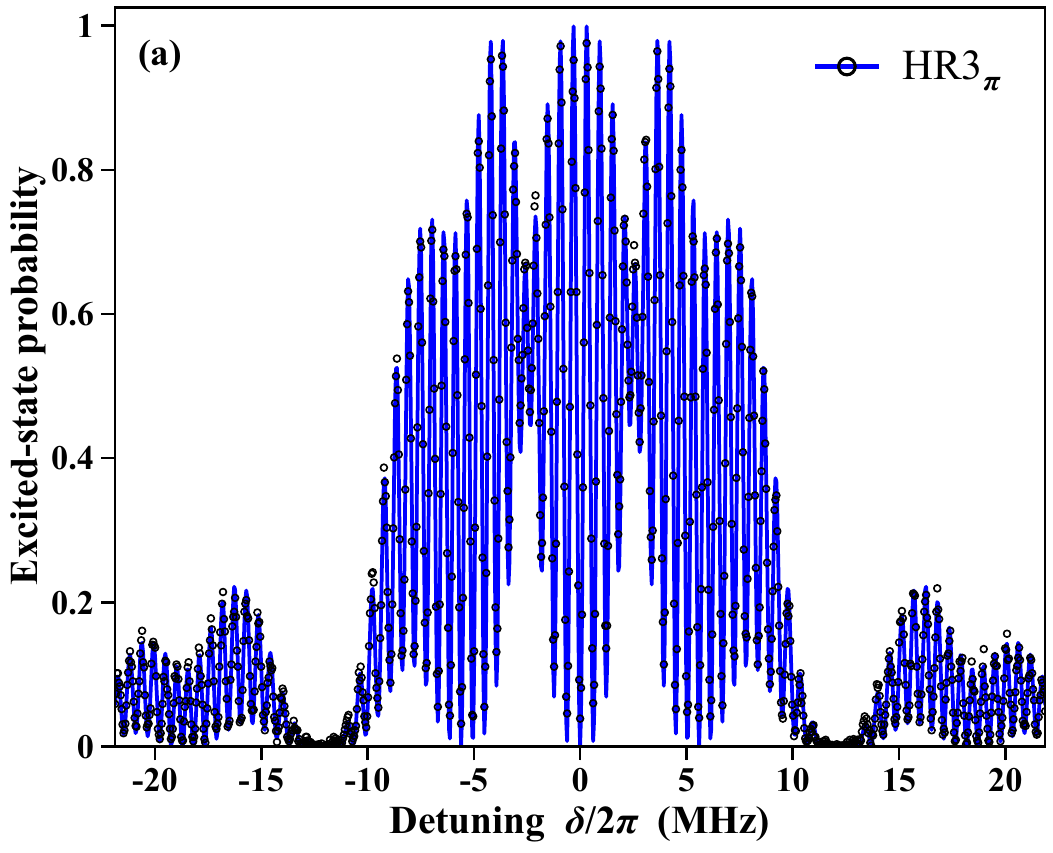}
\includegraphics[width=8cm, height=5.5cm]{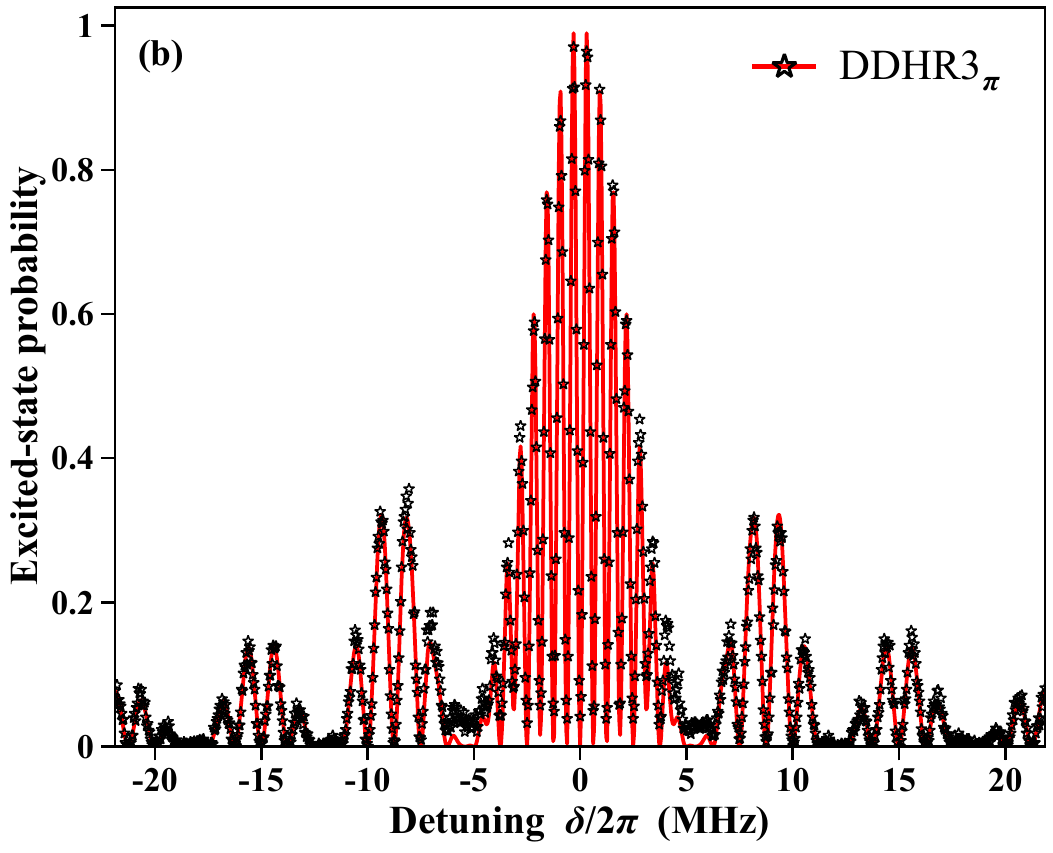}
\caption{Experimental quantum interferences  measured on NQCH superconducting quantum computer versus the detuning with 1001 points over $\delta/2\pi\in[-21.875,21.875]$~MHz. (a) HR3$_{\pi}$(circles) and (b) DDHR3$_{\pi}$(stars) with their simulations~(solid lines) for comparison. Pulse parameters are $\Omega_{l}=\pi/2\tau$ (l$=1,2,3$) with duration $\tau=80~\mathrm{ns}$ and free evolution time T$=1.6~\mathrm{\mu s}$. No residual light-shift and no laser probe frequency drift are assumed to be present.}
\label{fig:oscillations-NQCH}
\end{figure}

In clock laser spectroscopy, some methods have been explored to shield an optical qubit-clock transition from external influences by applying additional radio-frequency fields~\cite{Aharon:2019,Pelzer:2024,Akerman:2025}. Among optical techniques, hyper-Ramsey (HR) spectroscopy was proposed in 2010~\cite{Yudin:2010} and demonstrated experimentally in 2012~\cite{Huntemann:2012} to drastically reduce AC-Stark shifts induced by laser coupling to off-resonant atomic states and neighboring Zeeman sub-levels~\cite{Huntemann:2012,Yudin:2023}. A modified version of HR spectroscopy was proposed and demonstrated in 2016 within optical lattices probing ultra-narrow clock transitions of bosons eliminating probe-field-induced AC-Stark shifts by three orders of magnitude even with significant errors in shift compensation~\cite{Hobson:2016}. Single-ion optical clocks based on $^{171}$Yb$^{+}$~\cite{Huntemann:2016}, $^{40}$Ca$^{+}$~\cite{Huang:2022} and $^{176}$Lu$^{+}$~\cite{Zhiqiang:2023,Arnold:2026} have exploited HR spectroscopy reporting an impressive low uncertainty budget at or below the $10^{-18}$ relative level of accuracy. However, the asymmetric temporal position of the intermediate refocusing pulse in HR spectroscopy is required for the clock interferometer to be sensitive to the detuning of the probe beam providing the highest interference signal contrast. Consequently, the original HR protocol still suffers from residual clock frequency shifts related to decoherence and laser-probe-intensity-fluctuations~\cite{Tabatchikova:2013,Beloy:2018}. Robustness of HR spectroscopy and hybrid schemes has also been investigated in the presence of trapped-ion heating processes~\cite{Kuznetsov:2019} and in an optically dense medium~\cite{Barantsev:2020}. Furthermore, active mitigation protocols, such as auto-balanced Ramsey and generalized auto-balanced Ramsey techniques~\cite{Sanner:2018,Yudin:2018} can, in theory, eliminate the shift entirely.  However, these techniques require at least two error signals as discriminators to mitigate the residual shift.

\indent \textit{Theory-} Here, we introduce a new technique we call dynamically-decoupled hyper-Ramsey~(DDHR3$_{\pi}$) spectroscopy, that symmetrizes the position of the single phase-shifted refocusing pulse like a Hahn-echo to generate a strong dynamical decoupling effect~\cite{Zanardi:1999,Shaka:1987,Levitt:2008}. Our new spectroscopic interrogation scheme of an optical qubit is based on mathematical combinatorics within a Pascal's triangle presented in Fig.~\ref{fig:DD-HR-protocols} not only for realizing nuclear magnetic resonance NMR-like composite pulses~\cite{Hore:1983} but also to set single or multiple free-evolution zones between multiple phase-shifted refocusing pulses~\cite{Hennig:2001} (see also the End Matter section with a periodic table of composite pulse interferometric protocols). Dynamically-decoupled HR spectroscopy (DDHR), as an upgraded version of original HR spectroscopy, whose pulse parameters are respectively reported in Fig.~\ref{fig:DD-HR-protocols}(b) and Fig.~\ref{fig:DD-HR-protocols}(c), merges composite laser pulse Ramsey spectroscopy~\cite{Ramsey:1950,Zanon-Willette:2016,Zanon-Willette:2018,Zanon-Willette:2022-1,Zanon-Willette:2022-2} and Hahn-echoes~\cite{Hahn:1950,Carr:1954,Meiboom:1958} to preserve high robustness against probe-induced frequency shifts during the clock interrogation in presence of a finite laser probe linewidth. We introduce an intermediate rotary Hahn-echo pulse~\cite{Solomon:1959,Abragam:1961}, with a negative detuning~\cite{Vitanov:2015,Kyoseva:2019}, which compensates for low noise fluctuation of the trapping environment and slow frequency spectral drifts or potential frequency offsets of the probe in both free evolution zones~\cite{Chaneliere:2015,McCormick:2019,Sadzak:2022}. We successfully merge our three-pulse DDHR interferometric scheme with the classic approach that involves composite pulses around various axes to cancel error accumulation~\cite{Levitt:1986}. The real gain comes from the passive way in which the compensation is achieved - a simple detuning flip and repositioning of the refocusing pulse are sufficient. Only a single frequency scan of the quantum interferences as a discriminator is necessary to deliver most of the shift suppression. Laser probe parameters including light-shift correction and resonance frequency drifts or offsets are described in Fig.~\ref{fig:DD-HR-protocols}(b)-(c). The residual light-shift $\Delta_{LS}$ is induced by excitation pulses while a technical defect as a frequency offset or drift $\varepsilon_{P}$ of the laser probe can be present over the entire sequence of pulses.\\
\begin{figure}[t!!]
\centering
\includegraphics[width=8cm, height=6cm]{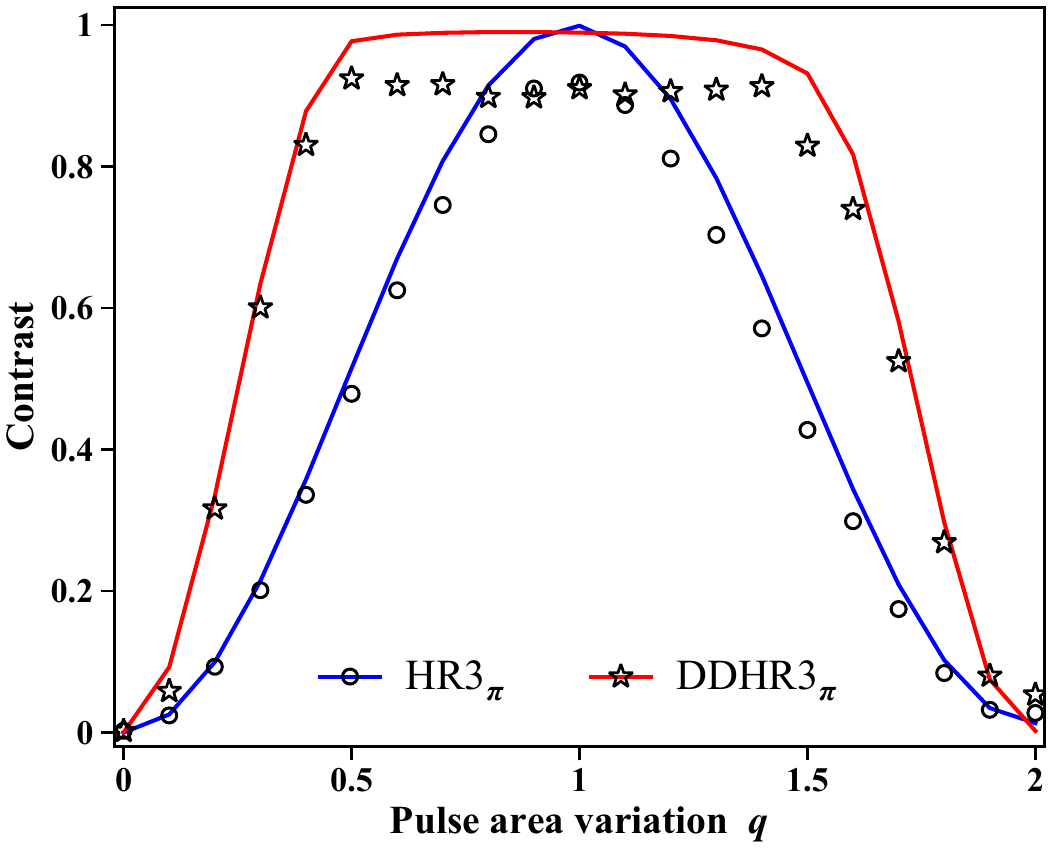}
\caption{Contrast of the superconducting NQCH qubit experiments for HR3$_{\pi}$~(blue line with circles) and DDHR3$_{\pi}$~(red line with stars) as functions of the pulse area $\Omega\tau=q\pi/2$. The experimental parameters match those in Fig.~\ref{fig:oscillations-NQCH}.}
\label{fig:contrast-NQCH}
\end{figure}

\indent \textit{Methods-} The HR and DDHR excitation probabilities are computed from the action of three laser pulses separated by one single or two free-evolution zones respectively. Ignoring decoherence, the SU(2) spinor dynamics of the $l$-th driving pulse ($l=1,2,3$) is the Cayley-Klein matrix~\cite{Zanon-Willette:2018}
\begin{equation}
\textup{M}(\theta^{\pm}_{\varphi_{l}}) \equiv \begin{pmatrix}
\cos\widetilde{\vartheta}_{l}e^{i\phi_{l}} & -i e^{-i\varphi_{l}}\sin\widetilde{\vartheta}_{l} \\
-i e^{i\varphi_{l}}\sin\widetilde{\vartheta}_{l} & \cos\widetilde{\vartheta}_{l}e^{-i\phi_{l}}
\end{pmatrix},
\label{eq:CK-matrix-body}
\end{equation}
where $\varphi_{l}$ is the pulse phase and the superscript $\pm$ labels the sign of the effective probe detuning during the pulse excitation,
\begin{equation}
\delta_{l}\rightarrow \pm\delta_{l}+\varepsilon_{P}, \qquad \delta_{l}=\delta-\Delta_{LS},
\label{eq:eff-detuning-body}
\end{equation}
so that the residual light shift $\Delta_{LS}$ enters during the pulse excitation while $\varepsilon_p$ acts as a probe-frequency drift or offset. The Cayley-Klein phase-angle parameters $\widetilde{\vartheta}_{l}$ and $\phi_{l}$ are:
\begin{equation}
	\begin{split}
\widetilde{\vartheta}_{l}=\arcsin\left[\frac{\Omega_{l}}{\omega_{l}}\sin\widetilde{\theta}_{l}\right], \hspace{0.25cm}\phi_{l}=\arctan\left[\frac{\delta_{l}}{\omega_{l}}\tan\widetilde{\theta}_{l}\right].
	\end{split}
	\label{phase-angles}
\end{equation}
The pulse area is $\widetilde{\theta}_{l}=\theta_{l}/2=\omega_{l}\tau_{l}/2$ with a generalized Rabi frequency denoted as $\omega_{l}=\sqrt{\delta_{l}^{2}+\Omega_{l}^{2}}$.
The residual light-shift $\Delta_{LS}$ and the Rabi frequency $\Omega_{l}$ can be treated as independent or dependent parameters, following the hyper-Ramsey analysis of~\cite{Beloy:2018}. The free-evolution zone is governed by
\begin{equation}
\textup{M}(\pm)\equiv \begin{pmatrix}
e^{i(\pm\delta+\varepsilon_{P})T/2} & 0 \\
0 & e^{-i(\pm\delta+\varepsilon_{P})T/2}
\end{pmatrix},
\label{eq:free-evolution-body}
\end{equation}
with $T$ the free-evolution time. HR3$_{\pi}$ and DDHR3$_\pi$ interferences are then the ordered products of these pulse excitation and free-evolution matrices,
\small{\begin{subequations}\label{eq:HR3-DDHR3-body}
\begin{align}
\textup{HR3}_{\pi}&\equiv\textup{M}(90_{0}^{\circ,+})\,\textup{M}(180_{\pi}^{\circ,+})\,\textup{M}(+)\,\textup{M}(90_{0}^{\circ,+}),\label{HR3}\\
&\equiv\textup{M}(90_{0}^{\circ,+})\,\textup{M}(+)\,\textup{M}(180_{\pi}^{\circ,+})\,\textup{M}(90_{0}^{\circ,+}),\label{HR3-2}\\
\textup{DDHR3}_{\pi}&\equiv\textup{M}(90_{0}^{\circ,+})\,\sqrt{\textup{M}(-)}\,\textup{M}(180_{\pi}^{\circ,-})\,\sqrt{\textup{M}(+)}\,\textup{M}(90_{0}^{\circ,+}),\label{DDHR3}
\end{align}
\end{subequations}}
acting on the initial ground state, the corresponding excitation probability being $P_{e}=|\langle e|\,\textup{HR3}_{\pi},\textup{DDHR3}_{\pi}\,|g\rangle|^{2}$. The clock frequency discriminator survives the echo because the refocusing pulse carries the opposite detuning sign introduced as $\textup{M}(-)$. Unlike a standard Hahn-echo, the central refocusing $180_{\pi}^{\circ,-}$ pulse does not cancel the clock detuning, which means the phase gets accumulated and allows the entire modified protocol to act as a frequency discriminator.

We demonstrate both protocols by mapping each Cayley-Klein matrix $\textup{M}(\theta^{\pm}_{\varphi_{l}})$ [Eq.~(\ref{eq:CK-matrix-body})] directly to the calibrated pulse set using an He-cooled qubit from the National Quantum Computing Hub (NQCH) at Singapore. Square pulses of duration $\tau_l$ drive the rotations $\widetilde{\vartheta}_{l}$, while virtual $z$-rotations~\cite{McKay:2017} drive the pulse phases $\varphi_{l}$. The processor plays each detuning point on one carrier at a frequency of $f_{q}+(\delta-\Delta_{LS}+\varepsilon_{P})/2\pi$ where $f_{q}$ is the experimental calibrated qubit frequency. Thus the two protocols inject any frequency shifts with the same mechanism. Each free-evolution zone is a real hardware delay, during which the qubit accumulates phase. HR3$_{\pi}$ decays with $T_{2}^{*}$ like a standard Ramsey sequence, while the echo of DDHR3$_{\pi}$ acts as a dynamical decoupling pulse and extends the decay limit to $T_{2}$~\cite{Ithier:2005,Bylander:2011}. We give the full pulse-level execution of $\textup{M}(\theta^{\pm}_{\varphi_{l}})$ and $\textup{M}(\pm)$ onto the native hardware set in the End Matter. As we will show in the next part of  the paper, a striking feature of DDHR3$_{\pi}$ is that it attains a much higher robustness to probe field-induced perturbations than HR3$_{\pi}$ using the exact same resources.\\

\begin{figure}[b!!]
\centering
\includegraphics[width=8cm, height=6cm]{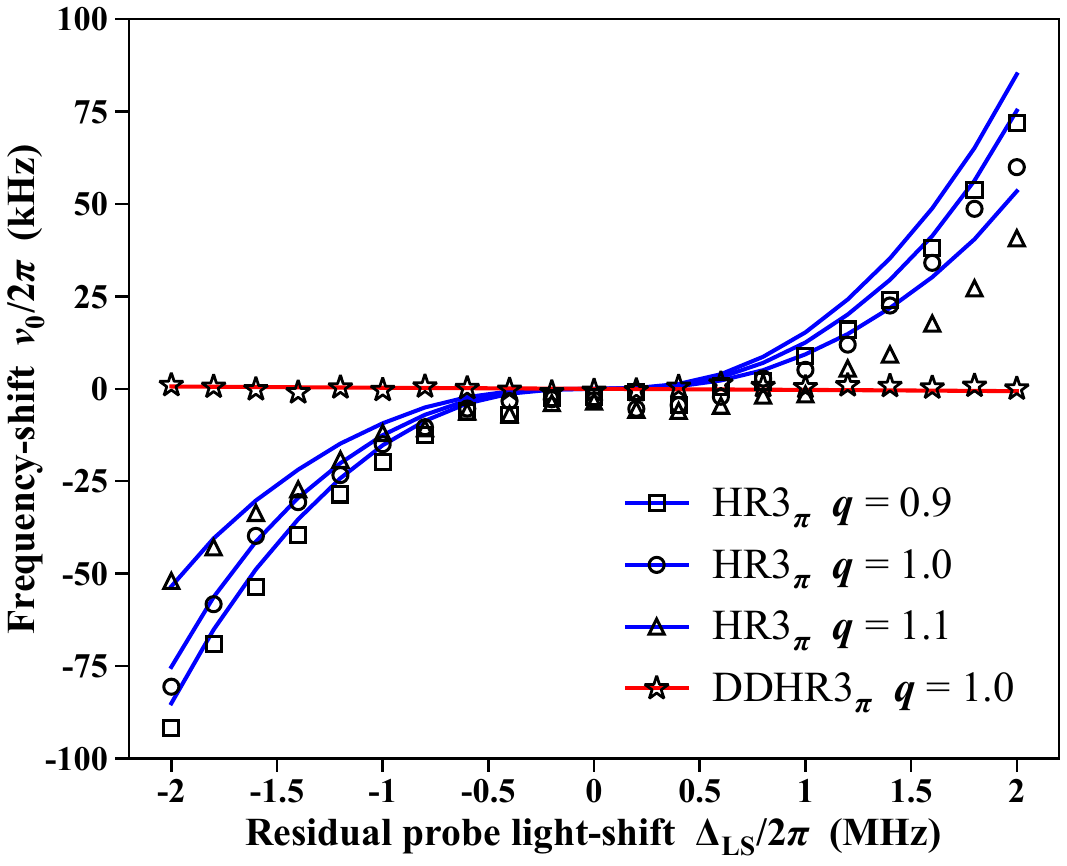}
\caption{Robustness of HR and DDHR protocols against residual probe-induced light-shift $\Delta_{LS}/2\pi$. Central-fringe shifts measured on the NQCH processor for HR3$_{\pi}$~(squares, circles and triangles) and DDHR3$_{\pi}$~(stars) with simulations (solid lines) for comparison. Microwave probe parameters include a potential variation of pulse areas by $\pm10\%$ ($\Omega\tau=q\frac{\pi}{2}$ where q$=0.9,1.0,1.1$). The experimental parameters match those in Fig.~\ref{fig:oscillations-NQCH}.}
\label{fig:robustness-light-shift}
\end{figure}

\indent \textit{Results-} Simulated and experimental HR3$_{\pi}$ and DDHR3$_{\pi}$ interferences versus a scan of the probe frequency detuning are reported in Fig.~\ref{fig:oscillations-NQCH} with the NQCH qubit processor. All experiments of the main text were also performed with the IQM \textit{Sirius} superconducting platform~(see the Supplemental Material). We measure an average per point difference~(residual) between the simulated and experimental oscillations to be $1.47\%$ for HR3$_{\pi}$ and $1.34\%$ for DDHR3$_{\pi}$. Our investigation starts by plotting the interference contrasts of the HR3$_{\pi}$ and DDHR3$_{\pi}$ signals versus a variation of Ramsey and refocusing pulse areas over the interrogation protocol, demonstrated on the NQCH processor in Fig.~\ref{fig:contrast-NQCH}. The NQCH experiments are performed in the microwave regime with a standard pulse duration $\tau=80~\mathrm{ns}$ and $T=1.6~\mathrm{\mu s}$. Our results are respectively shown as lines overlaid with markers: blue lines with black markers for HR3$_\pi$ and red lines with black stars for DDHR3$_\pi$. The continuous lines represent the simulations computed using the microwave-regime parameters, while the superimposed markers denote the measured experimental data. We observe that our DDHR3$_{\pi}$ protocol exhibits a broad optimal interference contrast plateau being insensitive to large variations of the pulse area. From the mean contrast residuals of the experimental demonstration~($4.54\%$ for HR3$_\pi$ and $6.2\%$ for DDHR3$_\pi$), it seems that even though DDHR3$_\pi$ has gained maximal contrast at half the temporal pulse area compared to HR3$_\pi$, it still performs worse. We note that this should not be interpreted as a shortcoming of the protocol itself, but a consequence of the low coherence times of the qubit, supported by the fact that both protocols reach a comparable maximum at optimal pulse area. Nevertheless, this is the first evidence of a very robust interrogation protocol realized by a temporal symmetrization of the single refocusing pulse between Ramsey pulses~\cite{Zanardi:1999,Vitanov:2015}.

\begin{figure}[t!!]
\centering
\includegraphics[width=8cm, height=6cm]{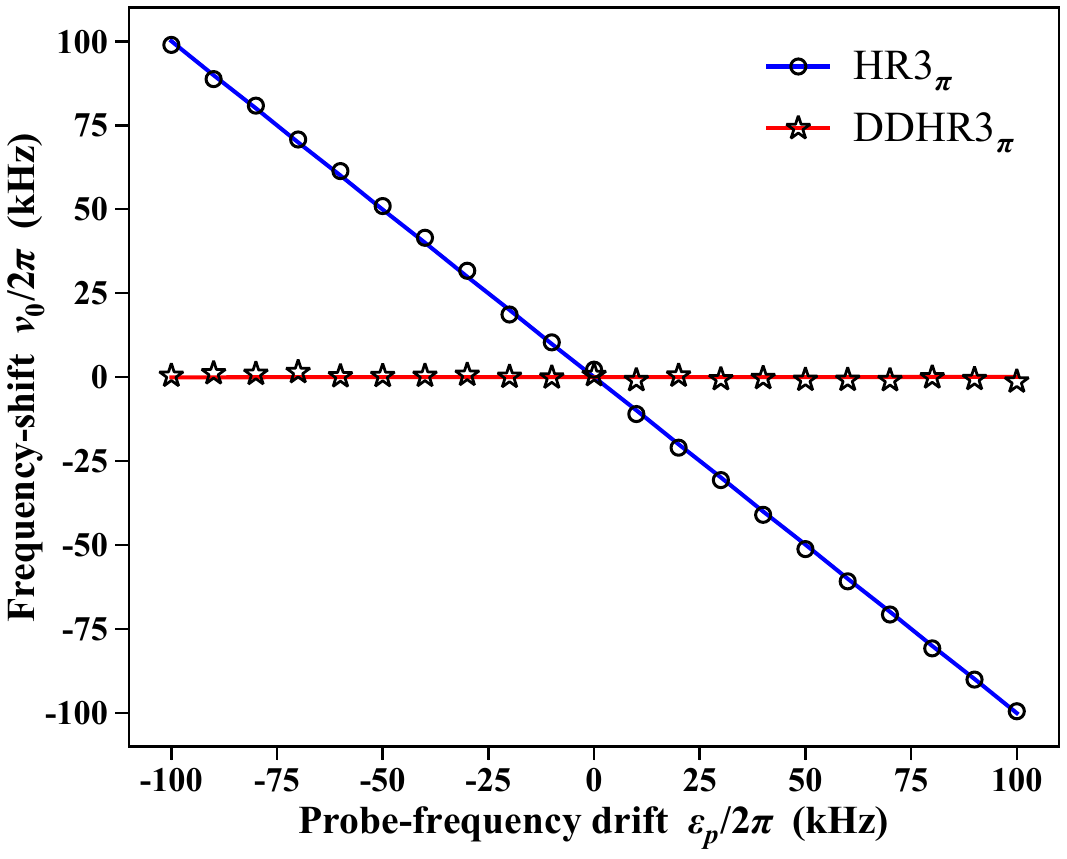}
\caption{Robustness of HR and DDHR protocols against a frequency drift of the probe in the microwave regime. Central-fringe shifts measured on NQCH for HR3$_{\pi}$ (circles) and DDHR3$_{\pi}$ (stars) with simulations (lines) for comparison. The experimental parameters match those in Fig.~\ref{fig:oscillations-NQCH}}
\label{fig:central-fringe-shift-eps}
\end{figure}

\indent We then evaluate the impact of probe-induced shifts of the transition line for HR3$_{\pi}$ and DDHR3$_{\pi}$ spectroscopy reported in Fig.~\ref{fig:robustness-light-shift} for the experimental demonstration in the microwave regime. For the experiments, the measured shift has mean residuals of $3.71$~kHz, $2.64$~kHz and $3.72$kHz for q=0.9/1.0/1.1 respectively, and $0.71$~kHz for DDHR3$_{\pi}$ at $q=1.0$. We can see the residual is about twice as large on the positive side~($8.3$ vs $3.6~\mathrm{kHz}$) for HR3$_\pi$, most likely caused by the state discrimination, qubit anharmonicity or a small pulse-amplitude calibration error. The HR3$_{\pi}$ frequency shift of the central fringe exhibits a nonlinear cubic dependence with the probe-induced light shift $\Delta_{LS}$~\cite{Yudin:2010,Huntemann:2012}, while the DDHR3$_{\pi}$ shift is completely suppressed by action of time reversal symmetry with a negative probe detuning of the refocusing pulse~\cite{Souza:2012-1}. We also have carefully checked that optical-clock simulations probing an ultra-narrow atomic or molecular transition with a pulse duration $\tau=10~\mathrm{ms}$ and a free-evolution time $T=200~\mathrm{ms}$ show the same predicted robustness. Indeed, the optical simulations operate at a different scale than the superconducting quantum computer experiments, where high error rates prevent us from measuring the effects at that same scale. Maintaining an invariant $\tau/T = 0.05$ ratio therefore ensures identical system dynamics.

We further deeply investigate, in the End Matter section for the clock regime, the decoherence impact due to a finite laser linewidth on the central fringe frequency shift, shown in Fig.~\ref{fig:decoherence-noise}(a) as a function of the residual probe-induced light-shift. As expected, the temporal symmetrization of the intermediate refocusing pulse provides a high robustness against residual AC-Stark shifts even in the presence of decoherence and can be even completely eliminated by applying dispersive error signals~\cite{Zanon-Willette:2016,Zanon-Willette:2018}.

\indent We then test our protocols against a quasi-static frequency offset or drift $\varepsilon_p$ of the central fringe frequency shift by averaging external low noise distortions over the entire interrogation protocol in Fig.~\ref{fig:central-fringe-shift-eps}, with a mean residual of $1.05$~kHz for HR3$_{\pi}$ and $0.65$~kHz for DDHR3$_{\pi}$. As expected, a linear dependence versus the probe frequency drift is established for HR3$_{\pi}$, which is strongly suppressed for DDHR3$_{\pi}$ due to the Hahn-echo configuration. The rotary Hahn-echo pulse reverses the time evolution of the qubit dynamics leading to an efficient compensation of the accumulated phase related to the spectral offset or drift of the probe laser.

At least, we have investigated the detrimental action of a small error in applying the relative phase-jump $\varphi=p\pi$ through the p parameter for two different protocols replacing the single DDHR3$_{\pi}$  rotary Hahn-echo refocusing pulse by a DDHR5$_{\pi}$ composite pulse following a Pascal binomial tree structure in the End Matter section. We have plotted the clock frequency shifts of these protocols versus the probe-induced light shift $\Delta_{LS}$ in Fig.~\ref{fig:phase-error}(a) and versus a quasi-static frequency drift $\varepsilon_p$ in Fig.~\ref{fig:phase-error}(b) for a range of refocusing-pulse phase-jump errors up to 10$\%$. Such a replacement is demonstrated to efficiently suppresses critical errors in phase-jumps.
In addition, we have investigated the effect of the fluctuation of the probe frequency offset on contrasts of HR3$_{\pi}$ and DDHR3$_{\pi}$ interferences (results shown in the Supplemental Material). As expected for quasi-static noise, a single Hahn-echo pulse within a three-pulse DDHR3$_{\pi}$ interrogation scheme is largely sufficient to remove detrimental variations of the probe parameters opening the possibility to access longer free evolution times by adding more refocusing pulses in harsh environments~\cite{Jones:2024}.
\begin{figure}[t!!]
\centering
\includegraphics[width=8cm, height=6cm]{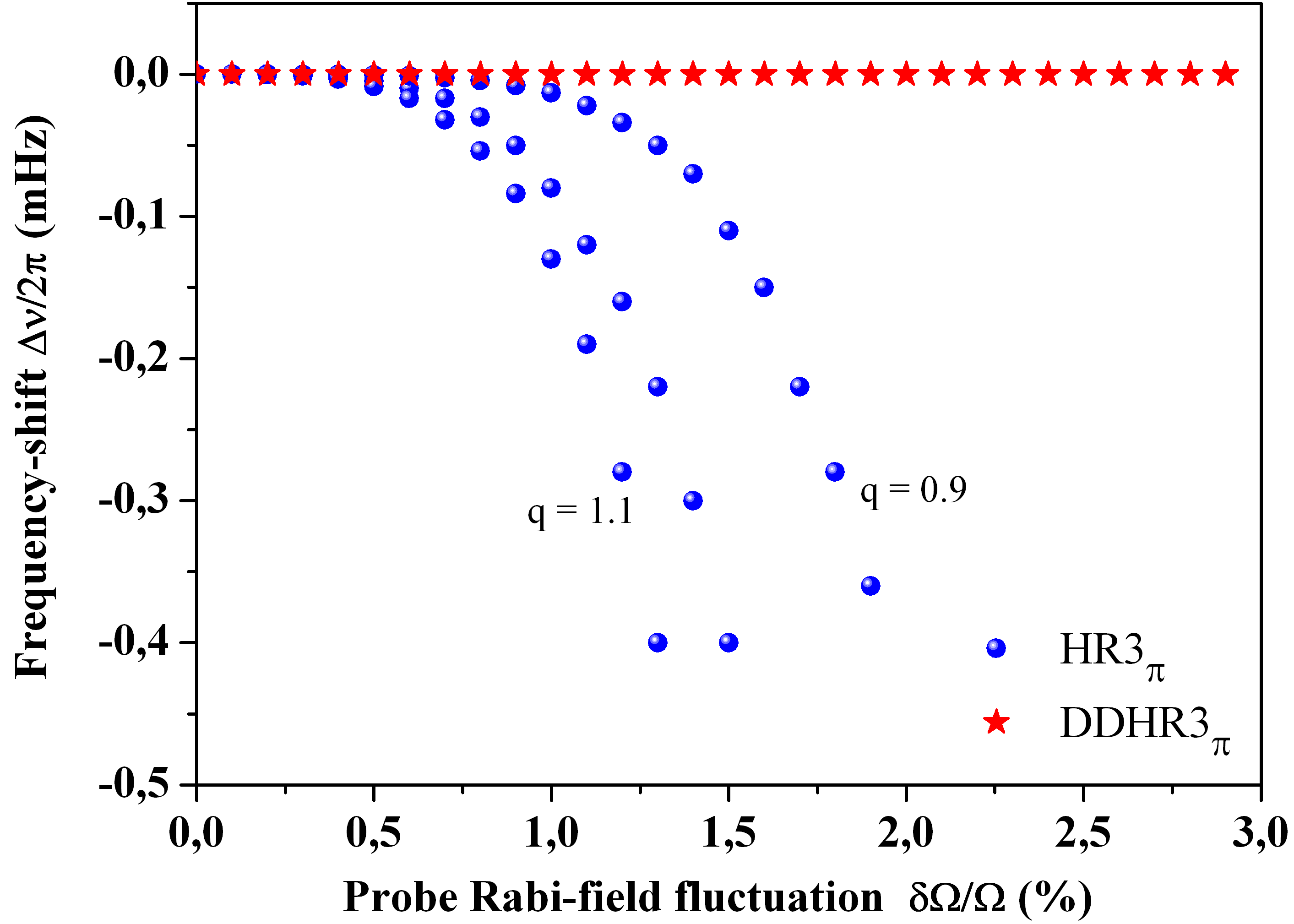}
\caption{Optical clock simulation of the robustness of HR3$_{\pi}$ and DDHR3$_{\pi}$ protocols with probe-laser-intensity fluctuations. Rabi field fluctuations are coupled to probe-induced light-shifts under a large pulse area variation by $\pm10\%$ ($\Omega\tau=q\frac{\pi}{2}$ where q$=0.9,1.0,1.1$) following the model in ref~\cite{Beloy:2018}.
}
\label{fig:field-fluctuation}
\end{figure}

We have finally theoretically investigated in Fig.~\ref{fig:field-fluctuation} the effect of a fluctuation of the Rabi laser probe field coupled to probe-induced light-shifts following~\cite{Beloy:2018} for the optical-clock regime. The temporal symmetrization of the intermediate refocusing pulse provides the highest robustness against residual AC-Stark shifts not only in presence of decoherence~\cite{Tabatchikova:2013,Zanon-Willette:2018} but also against probe intensity fluctuations~\cite{Beloy:2018}. The HR3$_{\pi}$ sequence (blue solid dots) is compromised with a clock frequency-shift acquiring a lower-order quartic dependence with Rabi field fluctuations. This is another demonstration of the high fault-tolerance of our DDHR3$_{\pi}$ interrogation protocol against intensity noise and decoherence. Due to the small magnitude of these shifts, we were unable to experimentally measure them on the NQCH quantum computer, as shot noise distorted the target signal.

\indent \textit{Conclusion-} DDHR spectroscopy is a fault-tolerant interrogation protocol that has been designed to outperform hyper-Ramsey spectroscopy against residual probe-induced frequency shifts under realistic experimental constraints, including probe intensity fluctuations, residual probe laser spectral drifts, decoherence related to probe laser with a finite linewidth, and technical pulse imperfections coupled to ambient electromagnetic noise. In particular, DDHR achieves a direct strong attenuation of sensitivity to probe intensity fluctuations. This eliminates the requirement for auxiliary discriminators or interleaved servo loops characteristic of active schemes, such as auto-balanced and generalized auto-balanced Ramsey spectroscopy~\cite{Sanner:2018,Yudin:2018}. Our single-refocusing-pulse DDHR3$_\pi$ protocol offering resilience against decoherence will be systematically enhanced by scaling the sequence with multiple palindromic Eulerian phase-shifted refocusing pulses~\cite{Viola:2003,Viola:2004}. Our DDHR interrogation scheme will offer, in the future, a natural generalization to the insertion of periodic, concatenated, and non-periodic dynamically decoupled pulse sequences \cite{Khodjasteh:2005,Uhrig:2007,Souza:2012-2,Suter:2016}, presenting a flexible framework for contrast optimization and elimination of probe-induced frequency shifts in presence of time-dependent noise~\cite{Kabytayev:2014,Uys:2009,Biercuk:2011,Ball:2015}.
These extended DDHR schemes will guarantee a complete mitigation of residual AC-Stark shifts, a minimization of sensitivity to probe spectral drifts and a preservation of the phase sensitivity of the quantum interference fringe with respect to the probe frequency detuning at the interferometer output.
The robustness of the interrogation protocol could be further refined by integrating optimal quantum control~\cite{Khaneja:2005}, advanced pulse shaping~\cite{Freeman:1998}, supercycle-based composite pulse decoupling~\cite{Levitt:1981,Kristiansen:2006}, and reinforcement learning-driven dynamical decoupling design~\cite{Marrder:2025} to tailor the spectroscopic method to specific environmental noise profiles and trapping potentials.

Our DDHR class of interrogation schemes is highly relevant to state-of-the-art quantum metrology platforms, including optical tweezer clocks \cite{Young:2020,Kaufman:2021}, Coulomb-crystallized multi-ion clocks \cite{Hausser:2025,Yu:2026}, multi-ion highly charged clocks~\cite{Kozlov:2018,Lyu:2025}, nuclear clocks employing trapped $^{229}$Th ions or thorium-doped solid-state systems~\cite{Peik:2021}, neutral-atom optical lattice clocks utilizing high-order multipolar transitions~\cite{Ishiyama:2024,Klusener:2024} and doubly forbidden bosonic transitions~\cite{Taichenachev:2006,Barber:2006,Baillard:2007,Zyskind:2026}. These architectures critically require robust interrogation strategies to sustain long-lived atomic coherence in the presence of spatial inhomogeneities in the electric or magnetic field, spectral drifts, and severe line-broadening mechanisms. Ultimately, DDHR spectroscopy promises to reinforce next-generation quantum sensors operating with ultracold atoms and molecules, facilitating high-precision searches for dark matter, gravitational wave detection, and tests of physics beyond the Standard Model in perturbed environments~\cite{Safronova:2018,DeMille:2024,Zaheer:2025,Norcia:2017,Schaffrath:2025}.
Molecular systems offer unique opportunities to advance tests of fundamental physics, thanks to their numerous degrees of freedom and a rich spectrum of nearly degenerate hyperfine and rovibrational energy levels. For instance, molecules are increasingly used to test fundamental symmetries~\cite{Roussy:2023,Eduardus:2026}, to track the spatiotemporal variation of fundamental constants~\cite{Shelkovnikov:2008,Barontini:2022}, and to search for dark matter~\cite{Barontini:2022,Gaul:2020}. Furthermore, their strong intramolecular interactions, inherent symmetries, and long-lived rotational and vibrational states make molecules highly promising platforms for quantum information processing~\cite{Tesch:2002}. Because many of these experiments rely on measuring resonance frequencies of molecular transitions, the critical role of improved spectroscopic tools for frequency metrology remains essential.
Boosting Ramsey interferometry via advanced dynamical decoupling is also expected to unlock unprecedented accuracy in antimatter spectroscopy~\cite{Comparat:2024}, thereby expanding the frontiers of fundamental particle and nuclear physics at a deeper level of precision~\cite{Kawasaki:2025}.

\indent T. Z-W thanks Jun Ye, Jonathan Jones, Ilya Kuprov and Emeric de Clercq for discussions. This work has been supported by QuantERA [QUARCKS], by ANR Grant No. ANR-26-QUA2-0012-01 and co-funded by the European Commission. T. Z-W and B. I. Thanks Rainer Dumke and Christoph Hufnagel from the National Quantum Computing Hub (NQCH) in Singapore and IQM quantum computers for providing access to computing resources. D. W thanks the Singaporean Minister of Education, Grant No. MOE-T2EP50223-0004, for financial support. B. I thanks Ivo Mihov for discussions. The data supporting the findings of this study, together with the pulse-compilation and simulation code, are available from the corresponding author upon reasonable request.

\begin{figure*}[t!!]
\centering
\resizebox{17cm}{!}{\includegraphics[angle=0]{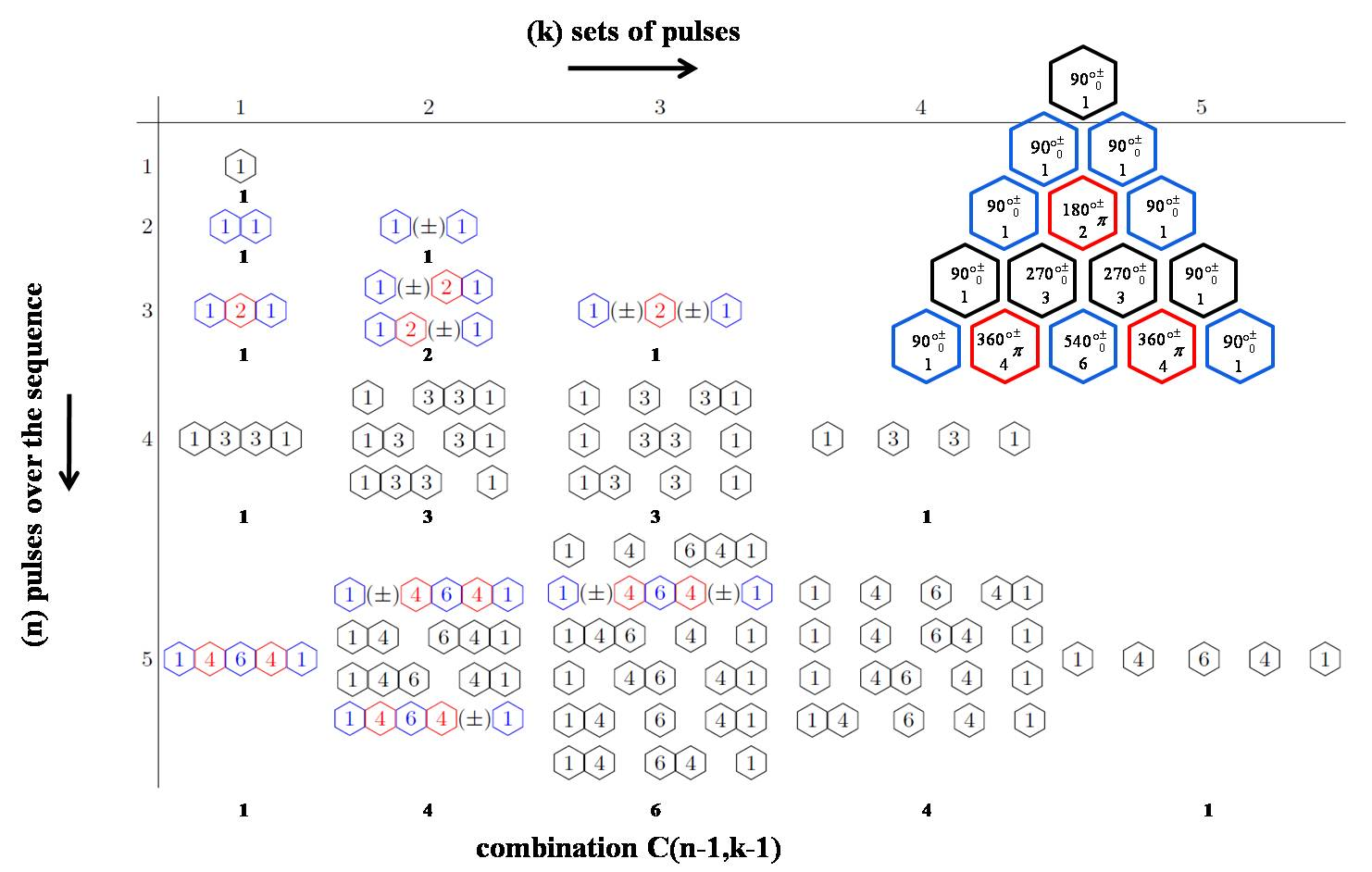}}
\caption{(top right) Pascal's triangle algorithm generating pulse protocols based on a multiple expansion of $90_{\varphi}^{\circ,\pm}$ Ramsey pulses. (center) Periodic table of composite pulse partitions $\{\textup{n,k}\}$ for laser spectroscopy, based on the composite-pulse construction of Eq.~(\ref{eq:binomial}), inserting ($\pm$) free evolution zones between pulses. The vertical axis is the number (n) of laser pulses and the horizontal axis is the number (k) of sets of pulses we are able to pack. We have highlighted a few sequences of pulses by red ($\varphi=\pi$) and blue ($\varphi_{}=0$) colored cells alternating phase jumps. The first column is a class of composite pulses used in NMR~\cite{Levitt:1986,Hore:1983}. The partition $\{2,2\}$ opens the second column of the periodic table with Ramsey spectroscopy based on a single free evolution zone encapsulated by two excitation pulses~\cite{Ramsey:1950}. The third column starts with a typical spin echo sequence~\cite{Carr:1954}.}
\label{fig:Pascal-triangle}
\end{figure*}

\centering
\onecolumngrid
\section*{End Matter}
\twocolumngrid
\justifying

\begin{figure*}[t!!]
\centering
\resizebox{8cm}{!}{\includegraphics[angle=0]{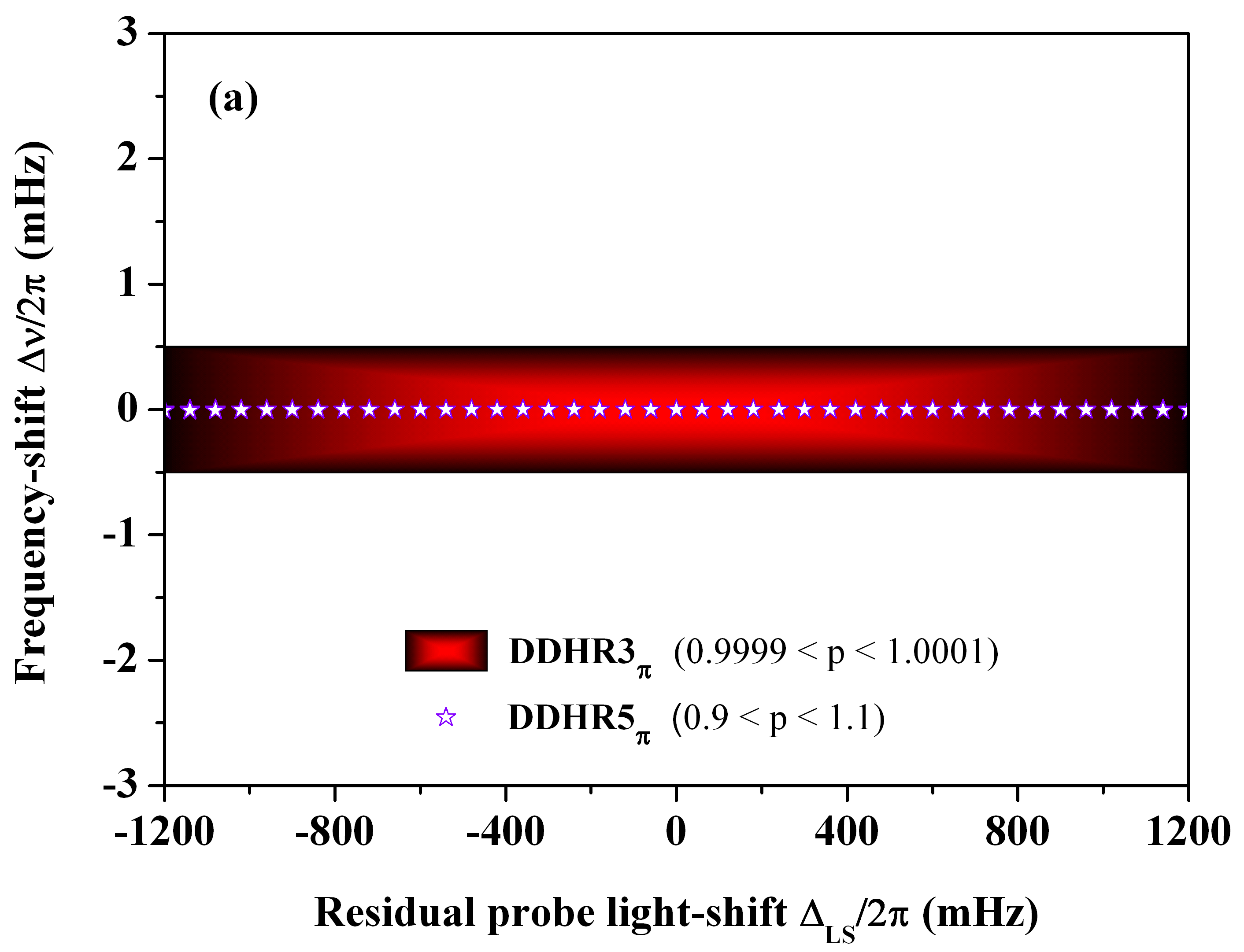}}\hspace{1cm}\resizebox{8cm}{!}{\includegraphics[angle=0]{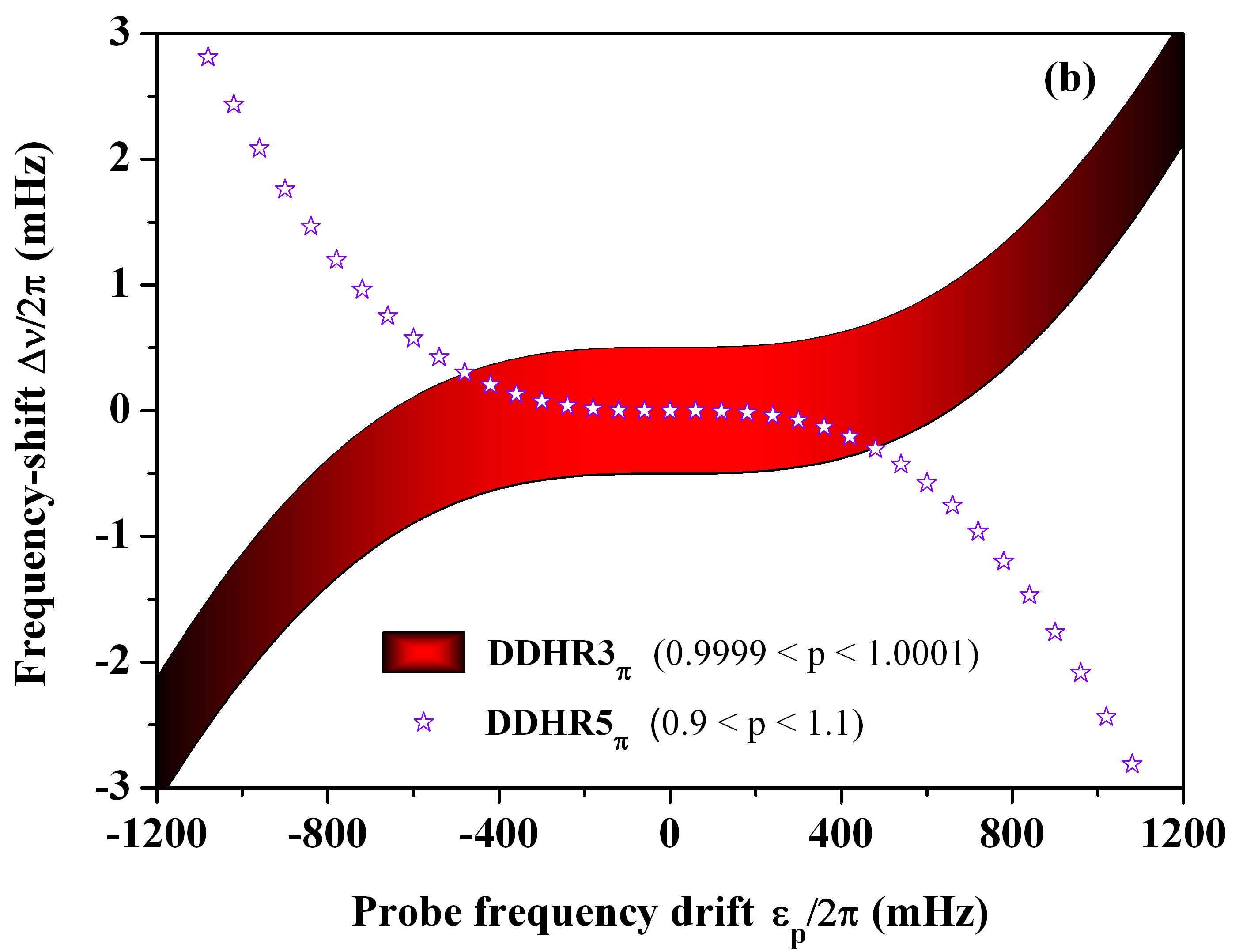}}
\caption{Optical clock simulations of the robustness of DDHR3$_{\pi}$ and DDHR5$_{\pi}$ sequences against a technical pulse defect like a potential error in phase-jumps. (a) Clock frequency shifts of DDHR3$_{\pi}$ (shaded region) and DDHR5$_{\pi}$ (open stars) versus residual probe-induced light-shift $\Delta_{LS}/2\pi$. (b) Clock frequency shifts of DDHR3$_{\pi}$ (shaded region) and DDHR5$_{\pi}$ (open stars) versus a laser probe frequency drift $\varepsilon_{P}/2\pi$. Employing a composite refocusing pulse with the DDHR5$_{\pi}$ sequence highly stabilizes the sensitivity to an error in the phase inversion.}
\label{fig:phase-error}
\end{figure*}

\indent-\textit{Experimental demonstration of HR and DDHR protocols}- For our experimental demonstration we use qubit~19 of the He cooled NQCH superconducting quantum processor. All runs use square pulses of duration $\tau=80$~ns at a calibrated Gaussian $\pi$-amplitude of $0.601$ (80~ns gate); the area theorem converts it into a square-pulse peak of $0.161$, which is then converted to an output voltage by a digital-to-analog converter~(DAC); the refocusing pulse is two times longer and the free-evolution time is $T=1.6~\mu$s. We recorded the three pulse areas $q=0.9,1.0,1.1$ and the DDHR3$_{\pi}$ point of the Stark sweep on a single calibration (10/08), so the four data sets are directly comparable and no recalibration occurred between them. The probe-drift sweep is the only exception and was done on 01/08. We apply SPAM readout for the measurements; at the interference calibration these are $0.39\%$ for $|g\rangle$ and $4.0\%$ for $|e\rangle$, with an average readout fidelity of $\approx98\%$. To construct the circuits we write the qubit as a two-level system, set $\hbar = 1$ and work in the rotating frame. The detuning $\delta$ is the difference between the drive of the microwave field and the measured resonant frequency of the qubit. The Hamiltonian reads
\begin{equation}
H(\delta,\Omega,\varphi) = \frac{1}{2} \begin{pmatrix}
-\delta & \Omega e^{-i\varphi} \\
\Omega e^{i\varphi} & \delta
\end{pmatrix},
\label{eq:ham}
\end{equation}
where $\Omega$ is the Rabi frequency and $\varphi$ is the drive phase.

\begin{figure*}[t!!]
\centering
\resizebox{8cm}{!}{\includegraphics[angle=0]{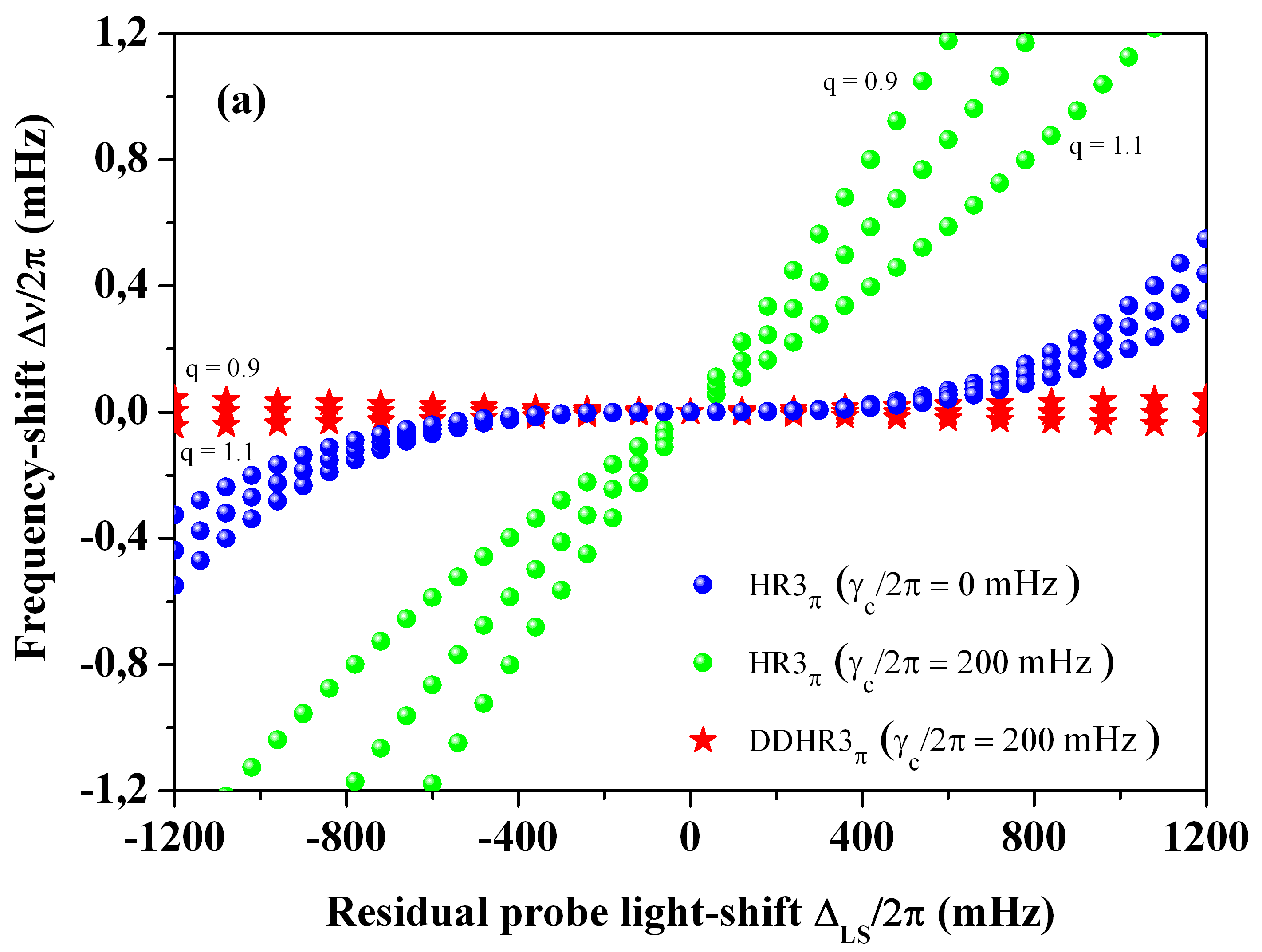}}\hspace{1cm}\resizebox{8cm}{!}{\includegraphics[angle=0]{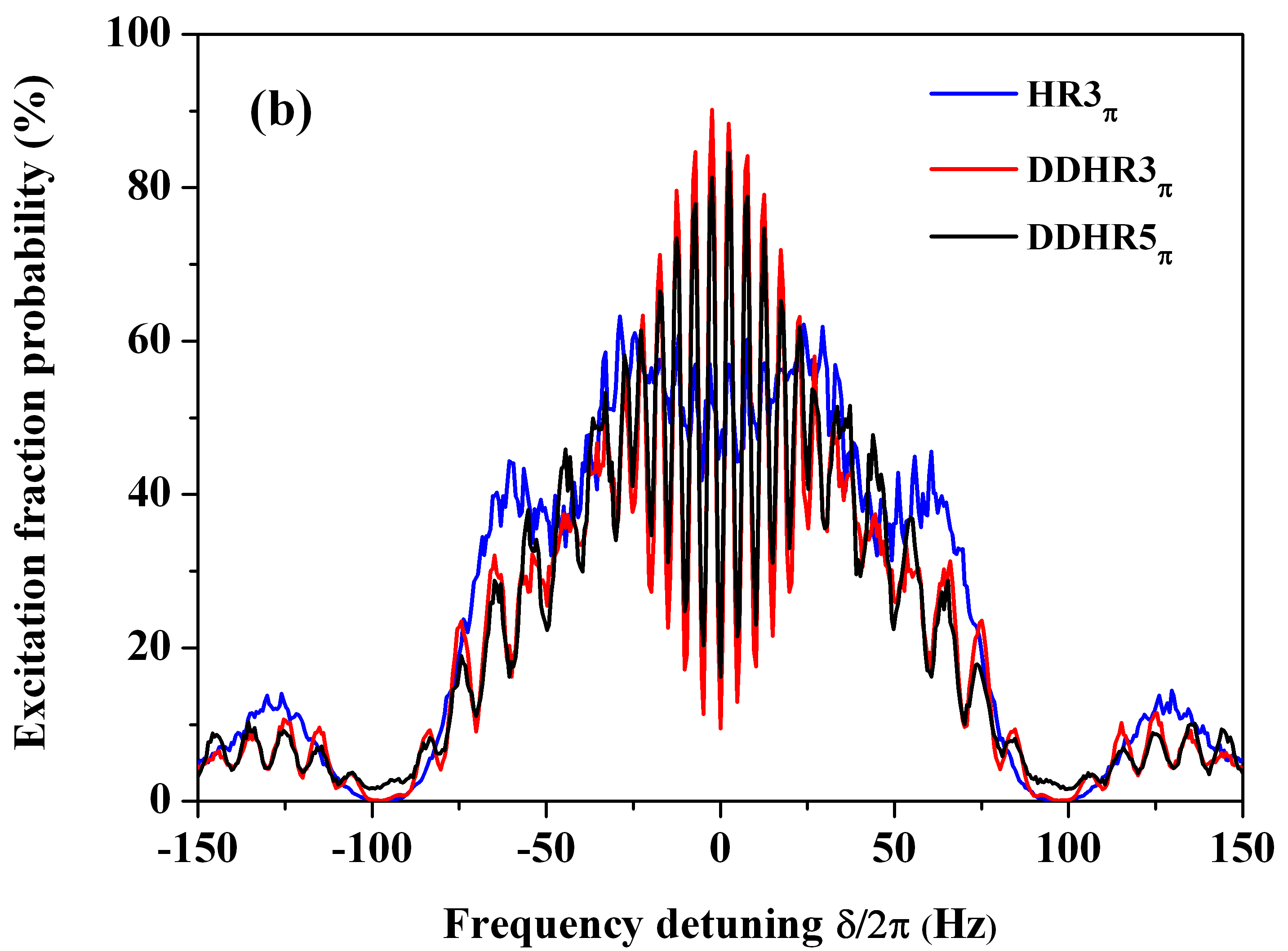}}
\caption{Optical clock simulations of the robustness of HR and DDHR protocols against residual probe-induced light-shift $\Delta_{LS}/2\pi$ in presence of decoherence and time-dependent noise. (a) HR3$_{\pi}$ and DDHR3$_{\pi}$ frequency shifts versus a residual probe-induced light-shift $\Delta_{LS}/2\pi$ in presence of decoherence $\gamma_{c}/2\pi=200$~mHz with a large variation of pulse areas by $\pm10\%$ ($\Omega\tau=q\frac{\pi}{2}$ where q$=0.9,1.0,1.1$). (b) Effect of a time-dependent stochastic noise on HR3$_{\pi}$, DDHR3$_{\pi}$ and composite DDHR5$_{\pi}$ interferences versus the probe detuning $\delta/2\pi$. Each plotted line is averaged over 200 runs from a Ornstein-Uhlenbeck stochastic noise distribution of amplitude strength $\sigma_{\mathrm{OU}}/\pi=3.5$. No AC-Stark shift and laser probe frequency offset or drift are assumed to be present.}
\label{fig:decoherence-noise}
\end{figure*}

For a constant drive we define the generalized Rabi frequency $\omega = \sqrt{\Omega^{2}+\delta^{2}}$ and the angle $\theta = \omega \tau/2$, where $\tau$ is the duration. The propagator then reads
\begin{equation}
U = \begin{pmatrix}
\cos\theta + i\dfrac{\delta}{\omega}\sin\theta & -i\dfrac{\Omega}{\omega}\,e^{-i\varphi}\sin\theta \\[6pt]
-i\dfrac{\Omega}{\omega}\,e^{i\varphi}\sin\theta & \cos\theta - i\dfrac{\delta}{\omega}\sin\theta
\end{pmatrix}.
\label{eq:prop}
\end{equation}
This propagator is equivalent to the Cayley-Klein parameterization for SU(2) spinor dynamics of Eq.~(\ref{eq:CK-matrix-body}), where the angle $\widetilde{\vartheta}$ and the accumulated phase $\phi$ are related to the physical pulse parameters by
\begin{equation}
\sin\widetilde{\vartheta} = \frac{\Omega}{\omega}\sin\theta \quad \text{and} \quad \tan\phi = \frac{\delta}{\omega}\tan\theta.
\label{eq:CK_mapping}
\end{equation}
We use this matrix to build every pulse, then concatenate the pulses in the specific sequence order given by equations (\ref{HR3}) and (\ref{DDHR3}). We initialize the qubit in the $\ket{0}$ state, execute the protocol, and measure the excited state population
\begin{equation}
P_{e} = |\,\langle 1|U|0\rangle\,|^{2},
\label{eq:pe}
\end{equation}
where $\ket{0}$ and $\ket{1}$ define the computational basis.  However, because superconducting architectures are weakly anharmonic, driving this transition can induce leakage into higher excited states, primarily $\ket{2}$ and $\ket{3}$. Therefore, the readout discriminates the ground state $\ket{0}$ from all other excited states.  Qubit frequencies and coherence times and the full details are given in the Supplemental Material.

\indent-\textit{Pascal's triangle algorithm and combinatorics}- Spectroscopic interrogation schemes of an optical qubit are supported by combinatorics of the Pascal's triangle not only for identifying NMR-like composite pulses but also to insert one or two free-evolution zones around a single or composite phase-shifted refocusing pulse.
In the context of optimized interrogation schemes, we use the HR3 scheme based on Eq.~(\ref{HR3}) as a standard protocol to compare spectroscopic performances with our DDHR3$_{\pi}$ protocol given by Eq.~\ref{DDHR3}. The number of $\{\textup{n,k}\}$ combinations using (n) pulses arranged into exactly (k) sets of pulses is given by the binomial coefficient:
\begin{equation}
\begin{split}
\textbf{C}(n-1,k-1)=\left( \begin{array}{c}
 n-1 \\
 k-1 \\
\end{array}\right)=\frac{(n-1)!}{(k-1)!(n-k)!}
\end{split}.
\label{eq:binomial}
\end{equation}
We report results into a periodic table of composite laser pulse partitions in Fig.~\ref{fig:Pascal-triangle}. The Pascal's triangle contains various sequences of refocusing laser pulses related to $\{\textup{n,k}\}$ partitions. By explicitly introducing signs of the laser probe detunings and specific values of phase jumps between laser fields into Pascal's partitions, we identify laser interrogation schemes associated to HR3$_{\pi}$ (HR5$_{\pi}$) protocols from $\{\textup{3,2}\}$ and $\{\textup{5,2}\}$ partitions and DDHR3$_{\pi}$ (DDHR5$_{\pi}$) protocols from $\{\textup{3,3}\}$ and $\{\textup{5,3}\}$ partitions.

\indent-\textit{DDHR composite pulses}- Our Pascal's triangle algorithm contains some sets of pulses that can be interpreted as composite pulses. Composite pulses are sequences of multiple single adjacent pulses with varying phases and durations designed to perform higher fidelity and robustness in experimental quantum control. From $\{\textup{3,2}\}$,$\{\textup{5,2}\}$ and $\{\textup{3,3}\}$,$\{\textup{5,3}\}$ partitions, we extract the following composite pulse:
\begin{equation}
\begin{split}
\textcolor{red}{\begin{tikzpicture}
  \foreach \a in {6}{
    \node[regular polygon, regular polygon sides=\a, draw, rotate=30,
     inner sep=0.02cm] at (\a*2,0) {\rotatebox{-30}{2}};}
\end{tikzpicture}}
\raisebox{1.5mm}{\(\;\equiv\)}\textcolor{red}{\begin{tikzpicture}
  \foreach \a in {6}{
    \node[regular polygon,regular polygon sides=\a, draw, rotate=30,
     inner sep=0.02cm] at (\a*2,0) {\rotatebox{-30}{4}};}
\end{tikzpicture}}\textcolor{blue}{\begin{tikzpicture}
  \foreach \a in {6}{
    \node[regular polygon, regular polygon sides=\a, draw, rotate=30,
     inner sep=0.02cm] at (\a*2,0) {\rotatebox{-30}{6}};}
\end{tikzpicture}}\textcolor{red}{\begin{tikzpicture}
  \foreach \a in {6}{
    \node[regular polygon, regular polygon sides=\a, draw, rotate=30,
     inner sep=0.02cm] at (\a*2,0) {\rotatebox{-30}{4}};}
\end{tikzpicture}}\label{eq:composite-DDHR}
\end{split}
\end{equation}
which can be translated into a palindromic or time-reversal sequence of three adjacent rotation pulses as:
\begin{equation}
\begin{split}
\textup{M}(180^{\circ,\pm}_{\pi})\equiv\textup{M}(360_{\pi}^{\circ,\pm})\textup{M}(540_{0}^{\circ,+})\textup{M}(360_{\pi}^{\circ,\pm})
\end{split}
\label{eq:composite-pulse}
\end{equation}
Here the integers denote pulse areas in units of $90^{\circ}$: the single refocusing pulse of area $2$ (a $180^{\circ}$ pulse) is replaced by the composite $4$-$6$-$4$ ($360^{\circ}$-$540^{\circ}$-$360^{\circ}$), the coefficients $4,6,4$ being a row fragment of the Pascal/binomial structure that generates the composite-pulse partitions. The composite pulse DDHR5$_{\pi}$ sequence can be computed by replacing the single intermediate refocusing phase-shifted Hahn-echo pulse with a composite pulse $\textup{M}(180_{\pi}^{\circ,\pm})\Leftrightarrow\textup{M}(360_{\pi}^{\circ,\pm})\textup{M}(540_{0}^{\circ,+})\textup{M}(360_{\pi}^{\circ,\pm})$ originally proposed for the HR5$_{\pi}$ scheme~\cite{Zanon-Willette:2022-1}.

We consider in Fig.~\ref{fig:phase-error} a detrimental error in controlling the relative phase-jump $\varphi=p\pi$ between laser pulses where $0.9\leq p\leq1.1$, plotting DDHR3$_{\pi}$ and DDHR5$_{\pi}$ clock frequency-shifts versus several laser probe parameters. By swapping the single rotary Hahn-echo pulse in DDHR3$_{\pi}$ by the composite pulse from Eq.~(\ref{eq:composite-pulse}), we report the modified robustness of the DDHR5$_{\pi}$ sequence versus residual AC-Stark shifts in Fig.~\ref{fig:phase-error}(a) and in front of a spectral frequency offset of the probe laser in Fig.~\ref{fig:phase-error}(b). The sensitivity of the DDHR protocol to an error in alternating phase jumps between $0$ and $\pi$ is removed by using a composite pulse that follow a Pascal binomial tree structure.

\indent-\textit{HR and DDHR protocols with decoherence and noise}- \indent The computation of HR and DDHR probabilities are based on the action of three laser pulses separated by one single or two free-evolution zones using SO(3) real matrices when a dephasing effect is present (see supplemental material). Here, we adopt a $3\times3$ real matrix $\textup{R}\left(\theta^{\pm}_{\varphi_{l}}\right)$ replacing the SU(2) $\textup{M}\left(\theta^{\pm}_{\varphi_{l}}\right)$ matrix for a l-th pulse which drives the qubit rotation on a Bloch sphere following a set of differential coupled equations as~\cite{Zanon-Willette:2018}:
\begin{equation}
\left\lbrace \begin{split}
\dot{\textup{U}_{l}}=&-\gamma_{c}~\textup{U}_{l} + \delta_{l}~\textup{V}_{l} - \Omega_{l}\sin\varphi_{l}~\textup{W}_{l},\\
\dot{\textup{V}_{l}}=&-\delta_{l}~\textup{U}_{l} -\gamma_{c}~\textup{V}_{l} + \Omega_{l}\cos\varphi_{l}~\textup{W}_{l},\\
\dot{\textup{W}_{l}}=&~\Omega_{l}\sin\varphi_{l}~\textup{U}_{l}-\Omega_{l}\cos\varphi_{l}~\textup{V}_{l}.\\
\end{split}\right.
\label{set-Bloch-DM}
\end{equation}
where $\gamma_{c}$ is a term of decoherence due to a finite laser linewidth. Optical coherence $\rho_{\textup{ge}}$ and population difference $\rho_{\textup{ee}}-\rho_{\textup{gg}}$ are related to density matrix elements by $\textup{U}\equiv\rho_{\textup{ge}}+\rho_{\textup{ge}}^{*}$, $\textup{V}\equiv i(\rho_{\textup{ge}}-\rho_{\textup{ge}}^{*})$ and $\textup{W}\equiv\rho_{\textup{ee}}-\rho_{\textup{gg}}$. Population conservation is given by the relation $\rho_{\textup{gg}}+\rho_{\textup{ee}}=1$. Complete solutions for HR3$_{\pi}$ and DDHR3$_{\pi}$ (DDHR5$_{\pi}$) quantum interferences that are driven by the previous set of equations including decoherence are provided in the Supplemental Material.

We evaluate the impact of decoherence on the central-fringe frequency shift as a function of the residual probe-induced light-shift $\Delta_{LS}$ under a large variation of pulse areas. The HR frequency-shift of the central fringe which exhibits a nonlinear cubic dependence with the probe-induced light-shift~\cite{Yudin:2010,Huntemann:2012} in absence of decoherence retrieves a linear dependance with the residual uncompensated part of the light-shift, as shown in Fig.~\ref{fig:decoherence-noise}(a). The DDHR frequency shift is strongly suppressed but still exhibits a very weak linear dependance with the residual light-shift that can be eliminated by using dispersive error signals~\cite{Zanon-Willette:2016,Zanon-Willette:2018}. We show the deleterious action of an external electromagnetic environment with a time-dependent stochastic noise in Fig.~\ref{fig:decoherence-noise}(b).

\clearpage
\onecolumngrid

\section*{Supplemental material:\\ Fault-tolerant hyper-Ramsey spectroscopy of optical clock transitions\\ with dynamical decoupling}

\section{Hardware execution on the NQCH superconducting qubit experiment}
\label{sec:nqch-methods}

The hardware plays every detuning point on one carrier. The carrier frequency is
\begin{equation}
f_{c}(\delta) = f_{q} + \frac{\delta - \Delta_{LS} + \varepsilon_p}{2\pi},
\label{eq:carrier}
\end{equation}
where $f_q$ is the calibrated qubit frequency listed in Table~\ref{tab:NQCH_calibrations}.
Both frequency offsets live in this carrier, meaning the light shift $\Delta_{LS}$ is present in the detuning of every pulse. The probe offset $\varepsilon_p$ shifts the pulse detuning and the free evolution detuning by the same amount. In the rest of this section we write $\delta_{p} \equiv \delta - \Delta_{LS}$ for the pulse-frame detuning. The probe offset $\varepsilon_p$ adds to every segment equally, so it cancels from every correction below, we omit it.

The DDHR3 refocusing $\pi$ pulse must be driven at the mirrored detuning $-\delta_{p}$. The carrier, however, already sits at $+\delta_{p}$ from the first $\pi/2$ pulse.  A second carrier for this one pulse is not an option, since the control system chains the phase of each waveform to the waveforms before it. In its own words, ''each waveform is multiplied by
the product of the outputs of the \texttt{phase\_update} for all preceding RF waveforms''~\cite{Keysight:QCS-SM}, and the channel property \texttt{absolute\_phase} makes each
waveform start at the end time of the previous
operation. One carrier keeps
the whole sequence on one phase clock so we keep the carrier
and embed the sign flip into the baseband envelope.

Every ramp-free pulse is an \texttt{RFWaveform} on the built-in envelope of the
run, \texttt{ConstantEnvelope} for the square pulses used here. The baseband ramp needs a phase that rotates within the pulse, so no built-in envelope can carry it. The middle $\pi$ of DDHR3 therefore runs as an \texttt{ArbitraryEnvelope}: a list of complex samples that carries $g(t)\,e^{iS_{\mathrm{ramp}}\delta_{p}t}$ sample by sample, where $g(t)$ is a square~(constant)
pulse envelope. The middle refocusing pulse carries the ramp $\exp[i\,S_{\mathrm{ramp}}\,\delta_{p}\,t]$ with $S_{\mathrm{ramp}}=-2$ and $t \in [0, 2\tau]$. This gives us the negative detuning $-\delta_{p}$, as Table~\ref{tab:detunings} requires. The refocusing $\pi$ also carries the drive phase $\varphi = \pi$, set as the \texttt{instantaneous\_phase} exactly as in HR3.  A \texttt{post\_phase} then unwinds the phase that the ramp leaves at the end of the pulse. Unlike the HR3 refocusing pulse, the DDHR3 one therefore carries two phase parameters: the \texttt{instantaneous\_phase} and the \texttt{post\_phase}. The HR3 pulse uses only the \texttt{instantaneous\_phase}.

The free time must run at $\delta$. The carrier frame, however, runs at $\delta_{p}$. This gap is a pure phase error: a delay has no drive, so the error acts only on the phase. We compensate with a virtual-$Z$ phase gate after every free delay. The control system plays this gate as a separate zero-duration waveform, \texttt{PhaseIncrement}, between the delay and the next pulse.  Once again, changing the carrier frequency for the free evolution periods mid experiment would cause the absolute phase reference to reset and desync from the rest of the experiment. Therefore, the delay still accumulates the full physical phase of the carrier, which is why the \texttt{PhaseIncrement} correction is needed.

The magnitudes are as follows: for HR3$_\pi$, the gate after the free evolution applies the phase $\Delta_{LS}\,T$. For DDHR3$_\pi$, after FE$_{1}$ we apply the phase $\Delta_{LS}\,T/2$, and after FE$_{2}$ we apply the phase $(\Delta_{LS} - 2\delta)\,T/2$. The ramp of the middle $\pi$ leaves the boundary phase $4\tau\delta_{p}$, which the post phase compensates.

At $\Delta_{LS} = \varepsilon_p = 0$, every repair term of HR3 and of the Ramsey control
vanishes. These sequences then need no phase gate at all. DDHR3 keeps three
repair terms: the ramp of the middle $\pi$, its post-phase, and the jump after
FE$_{2}$. These terms vanish only at $\delta = 0$.

The drive detuning, its negated counterpart, and the free detuning are therefore
\begin{align}
\delta_{\mathrm{dr}}& := \delta_p + \varepsilon_p, \qquad
\delta^{\mathrm{dr}} := -\delta_p + \varepsilon_p, \qquad \nonumber \\
\delta_{\mathrm{FE}} &:= \delta + \varepsilon_p, \qquad
\delta^{\mathrm{FE}} := -\delta + \varepsilon_{p}.
\label{eq:dets}
\end{align}
All experiments had 101 detuning points 1024 shots per point for their respective scans. The only exception was the contrast versus pulse area variation, which had 150 detuning points per scan.  Due to similar T$_2^*$/T$_2$ times, we are able to compare both protocols equally. Table~\ref{tab:detunings} lists the detuning of every segment, where HR3 and DDHR3 are the protocol frequencies, and NQCH is what we execute on the quantum computer.

\begin{table}[h]
\centering
\caption{Carrier frequency of each pulse segment.}
\label{tab:detunings}
\setlength{\tabcolsep}{7pt}
\begin{tabular}{ccccc}
\hline
\hline
Pulse & HR3 & DDHR3 & NQCH & IQM \\
\hline
$\pi/2$ & $\delta_{\mathrm{dr}}$ & $\delta_{\mathrm{dr}}$ & $\delta_{\mathrm{dr}}$ & $f_q$ \\
FE$_1$   & $\delta_{\mathrm{FE}}$ & $\delta_{\mathrm{FE}}$ & $\delta_{\mathrm{dr}}$ & $f_q$ \\
$\pi$    & $\delta_{\mathrm{dr}}$ & $\delta_{\mathrm{dr}}$ & $\delta_{\mathrm{dr}}$ & $f_q$ \\
FE$_2$   & $\delta_{\mathrm{FE}}$ & $\delta_{\mathrm{FE}}$ & $\delta_{\mathrm{dr}}$ & $f_q$ \\
\hline
\hline
\end{tabular}
\end{table}

\begin{table}[h]
\centering
\caption{Qubit-19 calibrations for each day of the experimental demonstration on the NQCH processor.}
\label{tab:NQCH_calibrations}
\setlength{\tabcolsep}{7pt}
\begin{tabular}{lccccc}
\hline
\hline
Date & $f_q$& $T_1$  & $T_2^*$  & $T_2$  &  RO err.  \\
 & $(\rm GHz)$ & $(\mu{\rm s})$ & $(\mu{\rm s})$ & $(\mu{\rm s})$ & $(\%)$ \\
\hline
10.08.2026 & 4.311885 & 30.2 & 23.9 & 25.1 & 2.3 \\
01.08.2026 & 4.311874 & 26.6 & 20.2 & 23.5 & 1.6 \\
\hline
\hline
\end{tabular}
\end{table}

\subsection{Pulse amplitude and the area theorem}
\label{sec:amplitude}

The calibration gives the amplitude of a Gaussian $\pi$ pulse. The protocol uses square pulses. The code converts the calibrated amplitude into the square-pulse peak with the area theorem. An edge pulse needs the pulse area $q\pi/2$. The peak drive amplitude is
\begin{equation}
A_{\mathrm{edge}} = \frac{q}{2}\,A_{\pi}\,\frac{\tau_{\mathrm{cal}}}{\tau}\,
                    \frac{\bar{g}_{\mathrm{cal}}}{\bar{g}_{\mathrm{proto}}},
\label{eq:amp}
\end{equation}
where $A_{\pi}$ is the calibrated Gaussian $\pi$ amplitude, $\tau_{\mathrm{cal}}$
is the calibration gate length, and $\bar{g}$ is the mean of the pulse envelope.
The calibration envelope is a truncated Gaussian so that it is zero at the gate edges
\begin{equation}
g_{\mathrm{cal}}(t) = \frac{e^{-(t-\tau/2)^{2}/(2\sigma_{\mathrm{g}}^{2})} - g_{0}}{1 - g_{0}},
\qquad \sigma_{\mathrm{g}} = \frac{\tau}{2 n_{\sigma}}, \quad g_{0} = e^{-n_{\sigma}^{2}/2}.
\label{eq:env}
\end{equation}
The mean of the envelope over the gate is
\begin{equation}
\bar{g}_{\mathrm{cal}} = \frac{1}{\tau}\int_{0}^{\tau} g_{\mathrm{cal}}(t)\,dt
        = \frac{\sqrt{\pi/2}\,\mathrm{erf}\!\left(n_{\sigma}/\sqrt{2}\right)/n_{\sigma}
                 - g_{0}}{1 - g_{0}}.
\label{eq:gmean}
\end{equation}
With $n_{\sigma} = 2$, Eq.~\eqref{eq:gmean} gives $\bar{g}_{\mathrm{cal}} = 0.535$. We measured the q19 calibration at $A_{\pi} = 0.601$~a.u. with $\tau_{\mathrm{cal}} = 80~\mathrm{ns}$.
For $q = 1$, Eq.~\eqref{eq:amp} gives the square-pulse peak
$A_{\mathrm{edge}} = 0.161$ a.u. The edge pulses run for
$80~\mathrm{ns}$, while the refocusing pulse ($\pi$) runs for $160~\mathrm{ns}$ at the
same amplitude. The runs at $q = 0.9$ and $q = 1.1$ use peaks of $0.145$ and $0.177$
a.u. This was done so that no peak leaves the DAC range $[-1,1]$.

\subsection{Readout correction}
\label{sec:readout}

The readout assigns each shot to the ground state or to the excited state. The
assignment makes two errors. Let $e_{0} = P(1|0)$ be the error that calls the
ground state excited. Let $e_{1} = P(0|1)$ be the error that calls the excited
state ground. The measured probability obeys the affine map
\begin{equation}
P_{e}^{\mathrm{meas}} = e_{0} + (1 - e_{0} - e_{1})\,P_{e}^{\mathrm{true}}.
\label{eq:spam}
\end{equation}
We invert this map to get the true probability,
$P_{e}^{\mathrm{true}} = (P_{e}^{\mathrm{meas}} - e_{0})/(1 - e_{0} - e_{1})$. This is commonly known as State Preparation and Measurement~(SPAM) error correction.

\subsection{Extraction of the fringe minimum $\mu_{0}$}
\label{sec:d0}

The light-shift and the probe-offset shift the central minimum $\mu_{0}$ of
the fringe. We locate the minimum with a template fit with the model
\begin{equation}
P_{e}(\delta) = A + B\,P_{\mathrm{sim}}(\delta - \Delta),
\label{eq:template}
\end{equation}
where $P_{\mathrm{sim}}$ is the simulated line shape. For a given shift $\Delta$, we scan $\Delta$ over the range $\pm 150~\mathrm{kHz}$, and reject a shift that turns the template upside down ($B < 0$). The measured minimum is
\begin{equation}
\delta_{0}^{\mathrm{meas}} = \delta_{0}^{\mathrm{sim}} + \Delta,
\label{eq:d0}
\end{equation}
where $\delta_{0}^{\mathrm{sim}}$ is the minimum of the simulated curve. We take that minimum with a parabola through the points near the bottom of the central dip~($\approx$ 0.1/T). No fit reached the $\pm 150~\mathrm{kHz}$ search bound, and the fit correlation stayed above 0.97.

\section{Methods for the IQM synthetic-evolution experiment}
\label{iqm-methods}

In this section we present the methods used to run the experiments on IQM \textit{Sirius} and how we process the data. The authors note that this is an ''emulation'' of the original protocols, aiming to find the theoretical maximum achievable with superconducting quantum computing architectures today. It is not a spectroscopic experiment aimed to provide evidence for the validity of the protocols, rather to show that cloud access to quantum computers-and creative use of some of their properties-allows one to check their theoretical predictions at minimal cost before committing to long lab experiments. The clearest example of this is the intensity field fluctuations mentioned in the main text, where even with the synthetic execution on IQM~(that does not suffer from almost any decoherence effects) we were unable to measure the shift accurately.

\subsection{Hardware parameters, and native gates}

IQM experiments were performed on qubit~1 of the IQM Sirius superconducting quantum processor, which uses fixed-frequency transmon qubits in a STAR24 topology. The PRX implementation used the default IQM calibrated \texttt{drag\_crf} waveform; no user-side DRAG re-tuning was applied. Ground state preparation at the start of each circuit and single-shot dispersive readout at the end were performed using the standard IQM control-stack routines with no readout-error mitigation applied at the user level. Calibration parameters for the two measurement days are summarized in Table~\ref{tab:iqm_calibration_app}. All figures shown in this Supplemental Material were acquired on 12.11.2025.

\begin{table}[b]
\centering
\caption{Calibration parameters for qubit~1 of the IQM \textit{Sirius} processor used in the synthetic-evolution experiments.}
\label{tab:iqm_calibration_app}
\begin{tabular}{lcccccc}
\hline
Date & $T_1$ & $T_2^*$ & $T_2$ & $f_q$ & PRX err. & RO err. \\
 & $(\mu{\rm s})$ & $(\mu{\rm s})$ & $(\mu{\rm s})$ & $(\rm GHz)$ & $(\%)$ & $(\%)$ \\
\hline
12.11.2025 & 26.43 & 7.37 & 20.86 & 4.7265 & 0.30 & 2.32 \\
\hline
\end{tabular}
\end{table}

The experimental parameters used throughout this section are
\begin{align}
\tau = 40~{\rm ns},
\qquad
\Omega/2\pi = \frac{1}{4\tau} = 6.25~{\rm MHz},
\qquad
T = 1~\mu{\rm s},
\end{align}
where $\tau$ is the pulse duration, $\Omega$ is the Rabi frequency, and $T$ is the free evolution time.

For each driven segment, we start from the Cayley-Klein propagator of the corresponding two-level pulse. The propagator matrix of the $l$-th pulse, $\theta^{\pm}_{\varphi_{l}}$ with $l=1,2,3$, driving SU(2) spinor dynamics reads~\cite{Zanon-Willette:2018-SM}
\begin{align}
\theta^{\pm}_{\varphi_{l}}
\equiv
\begin{pmatrix}
\cos\widetilde{\vartheta}_{l}e^{i\phi_{l}} & -i e^{-i\varphi_{l}}\sin\widetilde{\vartheta}_{l} \\
-i e^{i\varphi_{l}}\sin\widetilde{\vartheta}_{l} & \cos\widetilde{\vartheta}_{l}e^{-i\phi_{l}}
\end{pmatrix}.
\label{eq:Cayley-Klein-Matrix-1}
\end{align}
Here $\varphi_l$ is the programmed pulse phase. The superscript $\pm$ labels whether the segment is evaluated with positive or negative effective probe detuning,
\begin{align}
\delta_{l}\rightarrow \pm\delta_{l}+\varepsilon_p,
\qquad
\delta_{l}=\delta-\Delta_{\rm LS}.
\end{align}
Thus the residual light shift enters during the driven pulse segment, while $\varepsilon_p$ acts as a probe-frequency offset.

The Cayley--Klein phase angles are introduced as
\begin{align}
\widetilde{\vartheta}_{l}
=
\arcsin\!\left[
\frac{\Omega_l}{\omega_l}\sin\widetilde{\theta}_{l}
\right],
\qquad
\phi_l
=
\arctan\!\left[
\frac{\delta_{l}}{\omega_l}\tan\widetilde{\theta}_{l}
\right].
\label{phase-angles}
\end{align}
The pulse area is
\begin{align}
\widetilde{\theta}_{l}=\frac{\omega_l\tau_l}{2},
\qquad
\omega_l=\sqrt{\delta_{l}^{2}+\Omega_l^{2}},
\end{align}
where $\Omega_l$ is the Rabi frequency and $\omega_l$ is the generalized Rabi frequency.

The native single qubit pulse set consists of phased $x$-axis rotations, denoted PRX,
\begin{align}
R_{\varphi}(\theta)
=
\begin{pmatrix}
\cos\!\frac{\theta}{2} & -i e^{-i\varphi}\sin\!\frac{\theta}{2} \\
-i e^{i\varphi}\sin\!\frac{\theta}{2} & \cos\!\frac{\theta}{2}
\end{pmatrix},
\label{eq:IQM_PRX_app}
\end{align}
and virtual $z$-axis rotations,
\begin{align}
R_z(\phi)=
\begin{pmatrix}
e^{-i\phi/2} & 0 \\
0 & e^{i\phi/2}
\end{pmatrix}.
\label{eq:IQM_RZ_app}
\end{align}
Comparing Eq.~\eqref{eq:Cayley-Klein-Matrix-1} with the native PRX form gives the IQM implementation of each analytical pulse segment,
\begin{align}
\left[\widetilde{\vartheta}_{l}^{\pm}\right]_{\rm IQM}=R_z(-2\phi_l)\,R_{\varphi_l+\phi_l}\!\left(2\widetilde{\vartheta}_l\right).
\label{eq:IQM_gate_app}
\end{align}
The parameters $\widetilde{\vartheta}_l$, $\phi_l$, and $\varphi_l$ are computed from the analytical Cayley--Klein equations and passed directly to the IQM pulse parameters.

The free-evolution propagator $[\pm]$ obtained while switching the laser fields is
\begin{align}
[\pm]\equiv
\begin{pmatrix}
e^{i(\pm\delta+\varepsilon_p)T/2} & 0 \\
0 & e^{-i(\pm\delta+\varepsilon_p)T/2}
\end{pmatrix}.
\label{eq:Cayley-Klein-Matrix-2}
\end{align}
Using the IQM convention for virtual $R_z$ rotations in Eq.~\eqref{eq:IQM_RZ_app}, this Ramsey dark time phase is inserted as
\begin{align}
[\pm]_{\rm IQM} = R_z\!\left[(\mp\delta-\varepsilon_p)T\right].
\label{eq:IQM_evo_app}
\end{align}

Transition probabilities for the two protocols are obtained by multiplying the corresponding propagator matrices:
\begin{align}
{\rm HR3}_{\pi}
&\equiv
\left|
\left\langle e
\left|
90_{0}^{\circ+}180_{\pi}^{\circ+}[+]90_{0}^{\circ+}
\right|
g
\right\rangle
\right|^{2},
\label{HR-probability}
\\
{\rm DDHR3}_{\pi}
&\equiv
\left|
\left\langle e
\left|
90_{0}^{\circ+}[-]180_{\pi}^{\circ-}[+]90_{0}^{\circ+}
\right|
g
\right\rangle
\right|^{2}.
\label{DDHR-probability}
\end{align}
In the IQM circuits, the driven blocks are realized natively via Eq.~\eqref{eq:IQM_gate_app}, while the free-evolution blocks $[\pm]$ are inserted via Eq.~\eqref{eq:IQM_evo_app}. No physical idle delay is inserted during the dark intervals, so $T$ is an effective phase accumulation time, not a physical idle time of the qubit. Within this setting, the edge pulses have an area $\Omega\tau=\pi/2$, while the middle pulse has an area $\pi$. In the IQM implementation these are passed to the calibrated PRX gate as rotation angles in radians. Thus, varying $q$ corresponds to changing the effective drive amplitude, or equivalently the pulse area, at a fixed pulse duration.

\begin{figure*}[t!!]
\centering
\resizebox{8.33cm}{!}{\includegraphics[width=12cm,height=9.5cm]{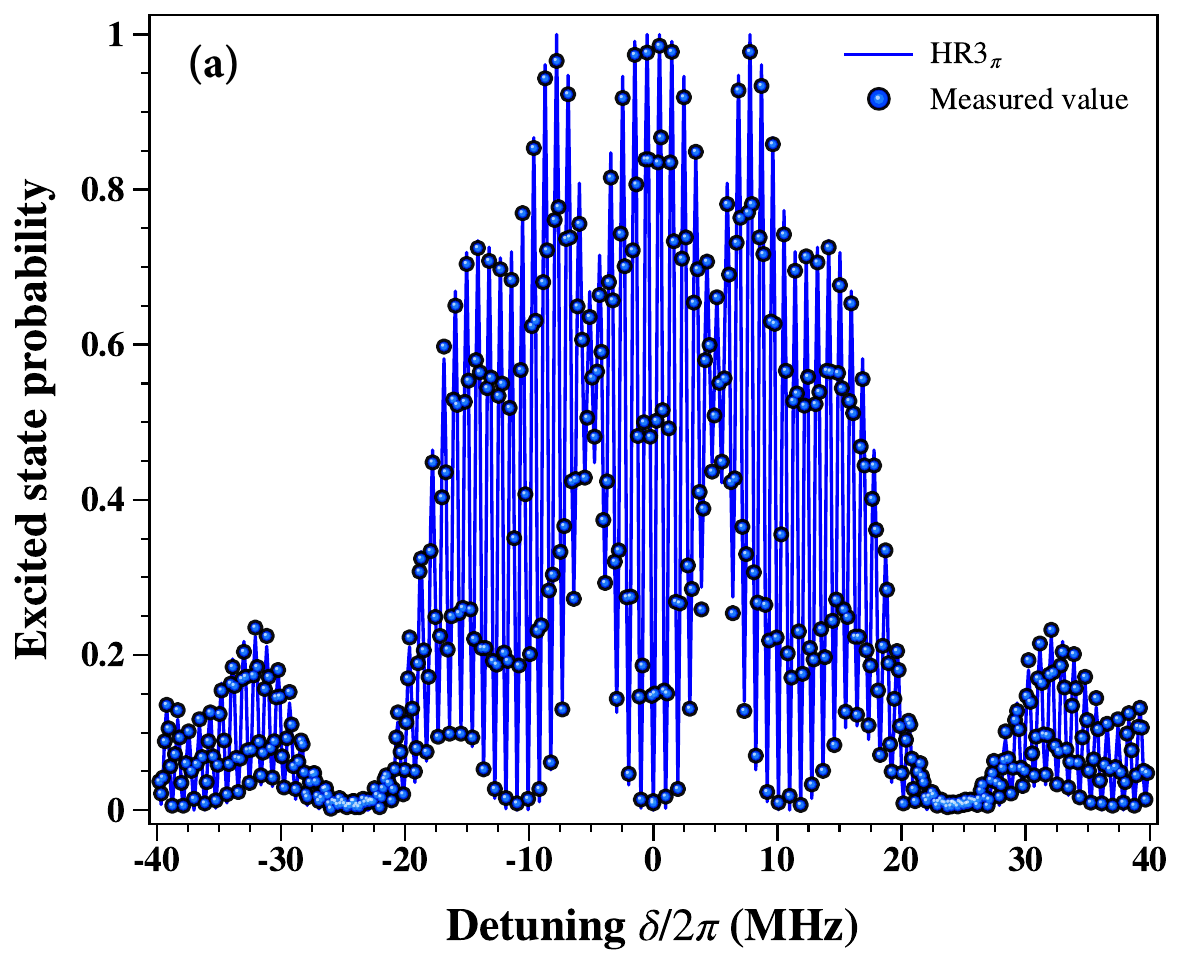}}\hspace{1cm}\resizebox{8.33cm}{!}{\includegraphics[width=12cm,height=9.5cm]{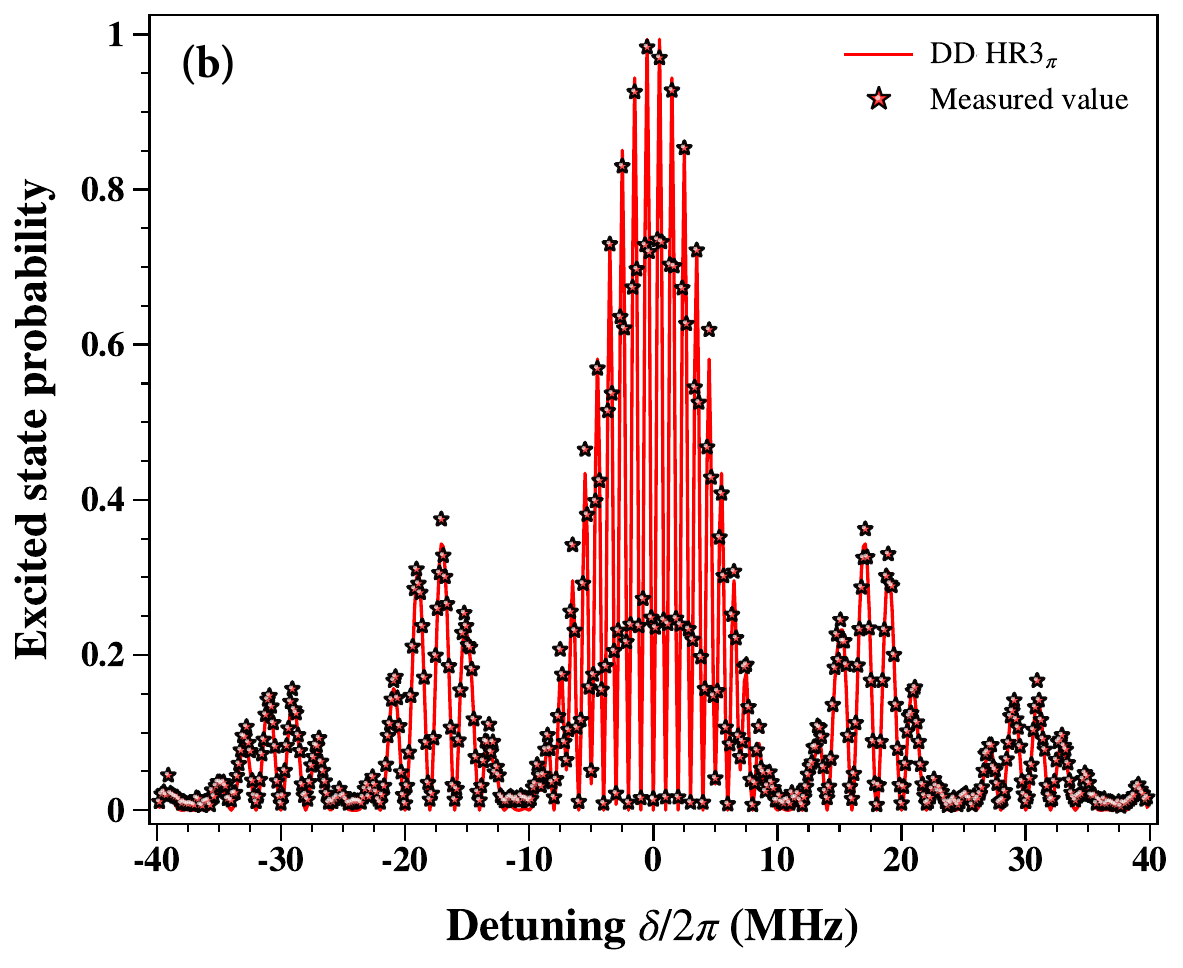}}
\caption{(a) Experimental HR3$_{\pi}$ and (b) experimental DDHR3$_{\pi}$ superconducting quantum interferences measured on IQM versus the laser probe detuning $\delta/2\pi\in[-40,40]$~MHz. Solid blue and red curves are provided by following the protocols of Eq.~(\ref{HR-probability}) and Eq.~(\ref{DDHR-probability}). The pulse parameters are $\Omega=\pi/2\tau$ with duration $\tau=40~\mathrm{ns}$ and free evolution time T$=1~\mathrm{\mu s}$, sampled with 501 points. No residual light-shift and no laser probe frequency drift are assumed to be present.}
\label{fig:interferences}
\end{figure*}

\subsection{Experimental demonstration of HR and DDHR protocols on the IQM processor}
\label{iqm-demonstration}

We demonstrate both protocols by mapping each Cayley-Klein matrix $\textup{M}(\theta^{\pm}_{\varphi_{l}})$ [Eq.~(\ref{eq:Cayley-Klein-Matrix-1})] directly to the calibrated pulse set of the IQM~\textit{Sirius} superconducting processor, where phased $x$-calibrated pulses (PRX, $R_{\varphi}(\theta)$~\cite{IQM:prx-SM}) and virtual $z$-rotations ($R_z(\phi)$~\cite{McKay:2017-SM,IQM:rz-SM}) drive the transitions at resonance.  The phase that the qubit would otherwise accumulate while evolving freely at a detuning $\delta$ is instead applied via a resonant $R_z$ rotation, with no physical idle~(free evolution) time, a technique we term \textit{synthetic evolution}.  While previous theoretical HR investigations usually assume zero quantum projection noise~\cite{Beloy:2018-SM}, with the advent of superconducting quantum processors and their error-free phase operation, we are able to execute the protocols while incorporating realistic errors from projection noise, readout, and pulse errors. Because of the different pulse structure in the two protocols, a real idle interval would inject this transmon's $T_2^*/T_2$ into the comparison and the qubit's coherence would be sampled unfairly on superconducting hardware~\cite{Ithier:2005-SM,Bylander:2011-SM}~(see Table~\ref{tab:iqm_calibration_app}). Removing the idle time isolates the coherent SU(2) dynamics that can be transferred across qubit platforms and allows for a more fair comparison of the two protocols, as neither protocol accumulates decoherence during the idle time. This allows us to use a minimal number of shots per experiment reducing execution time significantly.  A striking feature of DDHR3$_{\pi}$ is that it attains a high robustness to field-induced perturbations using the exact same resources as HR3$_{\pi}$. Simulated and experimental HR3$_\pi$ and DDHR$_\pi$ interferences versus a scan of the probe frequency detuning are reported in Fig.~\ref{fig:interferences}(a)-(b) from the IQM superconducting device; the corresponding interferences from the NQCH processor are reported in the main text; the analytical solutions are shown in Fig.~\ref{fig:SM-interferences}. This demonstrates the universality of our new interrogation protocol on two different experimental quantum computing platforms based on artificial atoms. Experimental data points are shown without error bars, we have reported the finite shot statistics in (Sec.~\ref{sec:iqm_shotnoise}).

\begin{figure}[t]
\centering
\includegraphics[width=8cm, height=6cm]{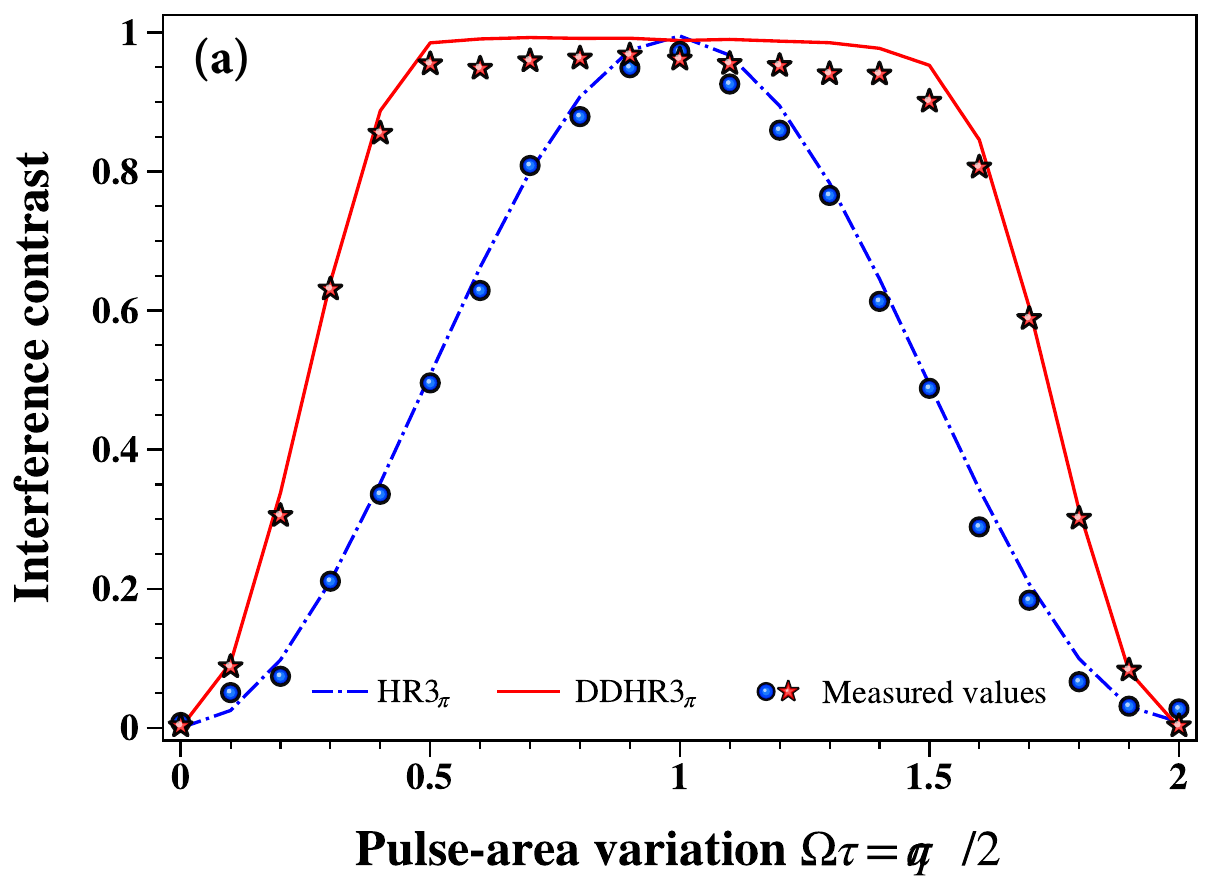}\hspace{1cm}\includegraphics[width=8cm, height=6cm]{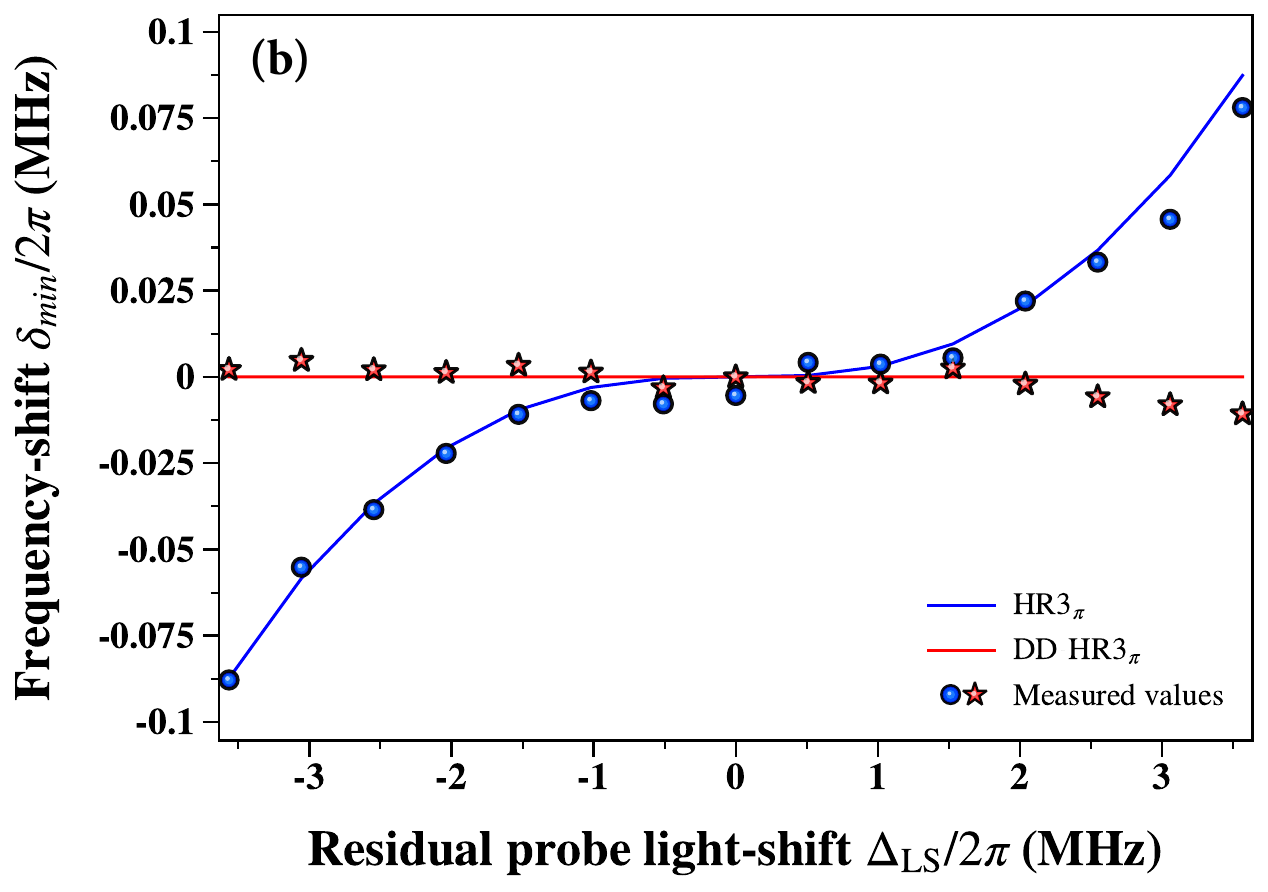}
\caption{(a) Excited state probability of the superconducting IQM qubit experiments for HR3$_{\pi}$~(blue circles) and DDHR3$_{\pi}$~(red stars) as functions of the pulse area, with simulations (solid red and blue lines) for comparison. (b) Central-fringe shifts of HR3$_{\pi}$ (blue circles) and DDHR3$_{\pi}$ (red stars) measured on IQM versus the residual probe-induced light shift $\Delta_{LS}/2\pi$, with simulations (lines) for comparison. Laser parameters are the pulse duration $\tau=40$~ns with a fixed free evolution time T$=1$~$\mu$s.}
\label{fig:IQM-fig3}
\label{fig:iqm-light-shift}
\end{figure}

We start by plotting the interference contrasts of the HR3$_{\pi}$ and DDHR3$_{\pi}$ signals versus a pulse area variation, demonstrated on IQM Sirius in Fig.~\ref{fig:IQM-fig3}(a). The IQM experiments are performed in the microwave regime with a standard pulse duration $\tau=40~\mathrm{ns}$ and $T=1~\mathrm{\mu s}$ and confirm the error scaling of the theoretical predictions. We observe that our DDHR3$_{\pi}$ protocol exhibits a broad optimal interference contrast plateau even under large variations of the pulse area.
We compare a numerical evaluation of excitation probabilities based on Eq.~(\ref{HR-probability}) and Eq.~(\ref{DDHR-probability}) (blue and red solid lines) with experimental data (blue circles and red stars) recorded on the \textit{IQM Sirius} superconducting qubit, with average residuals of 1.1086\% for HR3$_\pi$ and 1.1093\% for DDHR3$_\pi$.
The residual light-shift scan used 21 evenly spaced values over $\Delta_{\rm LS}/2\pi\in[-5.09,5.09]~{\rm MHz}$. For each value, the central minimum located from the simulation was experimentally sampled with 20 detuning points also uniformly distributed throughout the $\pm 0.318$~MHz local window and $\delta_{\min}$ was extracted using Eq.~\eqref{eq:reported_minimum_app}. The 3 lowest and 3 highest $\Delta_{\rm LS}$ values of were excluded from the final analysis because the fringe was sufficiently distorted that a single cosine model could not describe it well; this leaves 15 retained $\Delta_{\rm LS}$ values in the reported scan. The measured central-fringe shifts versus $\Delta_{\rm LS}$ are shown in Fig.~\ref{fig:iqm-light-shift}(b).

\begin{figure}[t]
\centering
\includegraphics[width=8cm, height=6cm]{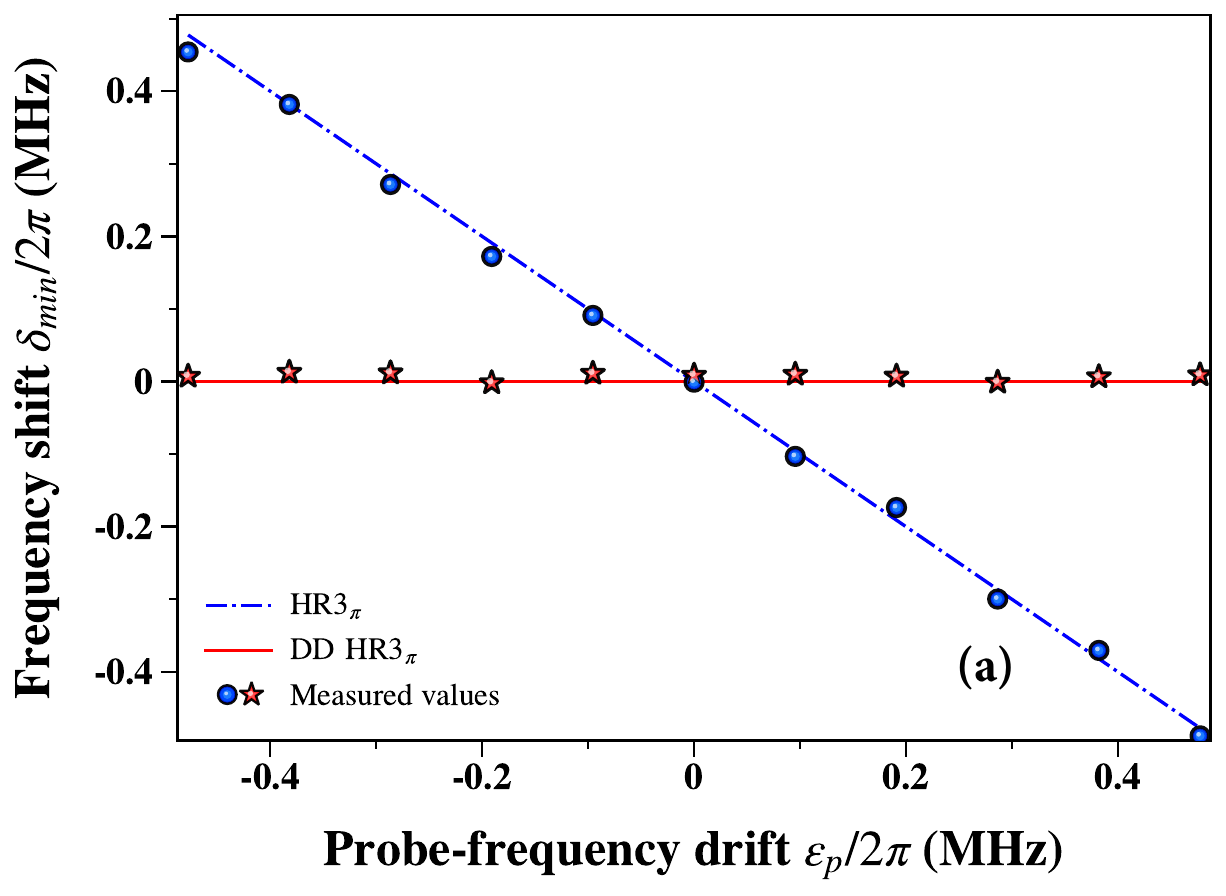}\hspace{1cm}\includegraphics[width=8cm, height=6cm]{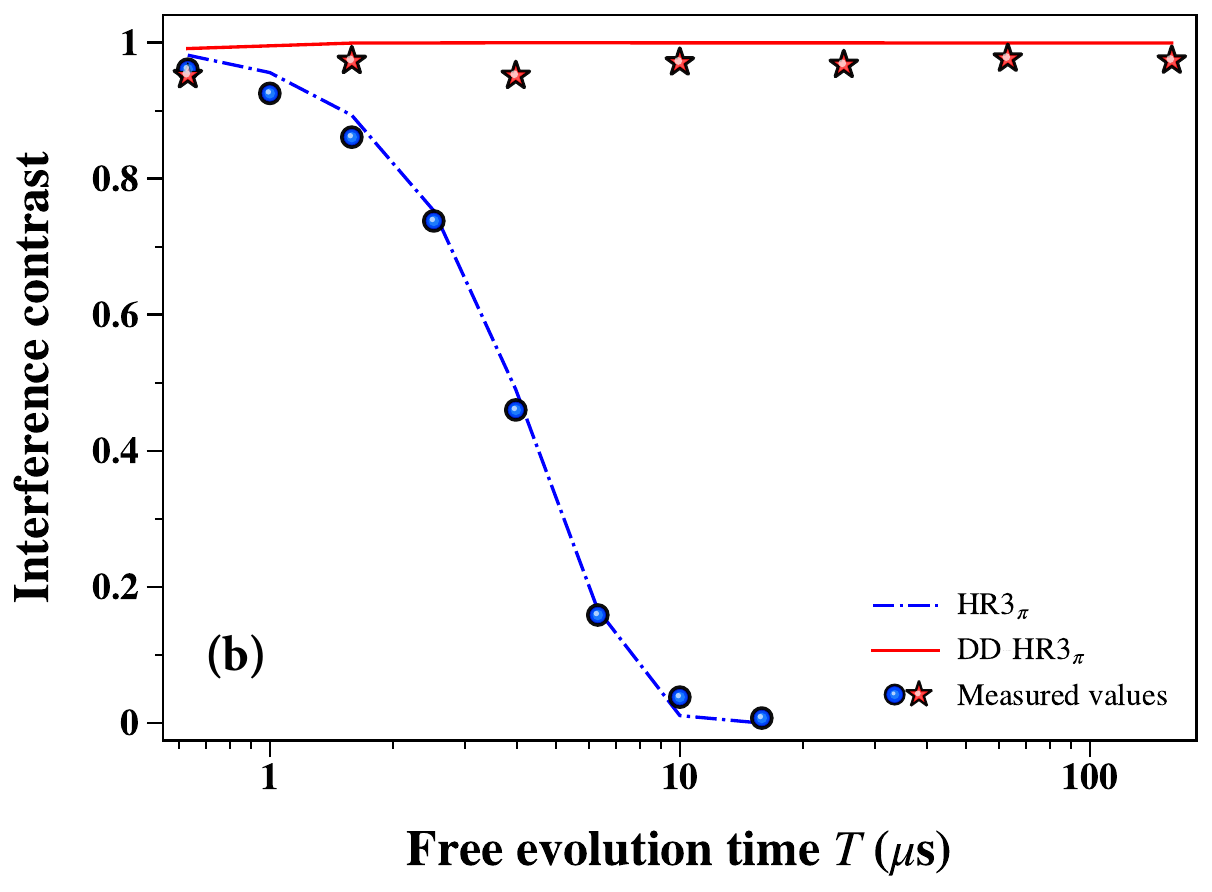}
\caption{(a) Central-fringe shifts of HR3$_{\pi}$ (circles) and DDHR3$_{\pi}$ (stars) measured on IQM versus the probe-frequency offset $\varepsilon_{p}/2\pi$, with simulations (lines) for comparison. All data correspond to a fixed pulse area ($\Omega\tau=q\frac{\pi}{2}$ with $q=1.0$). (b) Experimental measurement on IQM of HR3$_{\pi}$ and DDHR3$_{\pi}$ fringe contrasts versus the free evolution time.}
\label{fig:iqm-pfo}
\label{fig:quasi-static-noise-IQM}
\end{figure}

\begin{figure*}[t]
\centering
\includegraphics[width=8cm, height=6cm]{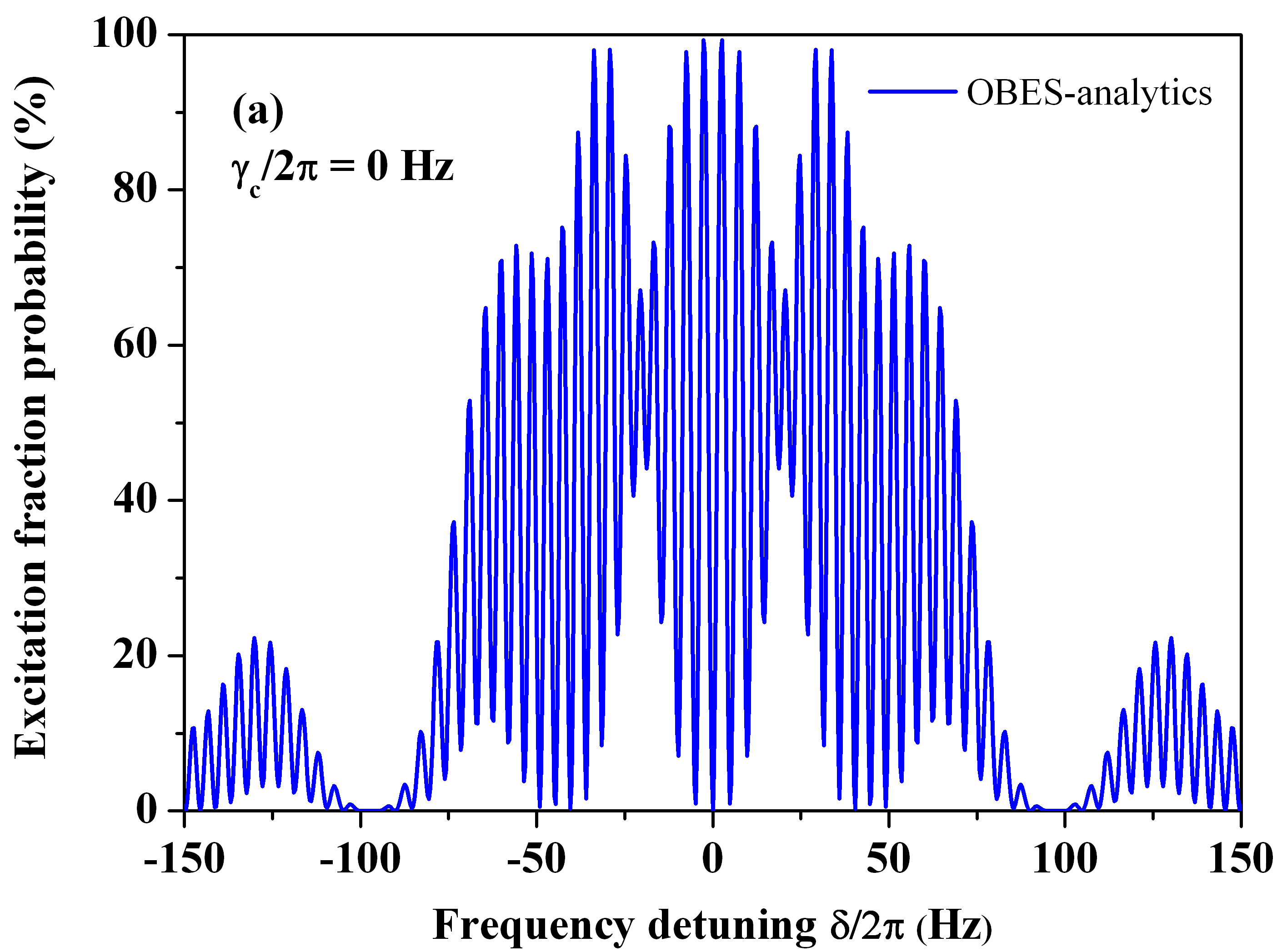}\hspace{1cm}\includegraphics[width=8cm, height=6cm]{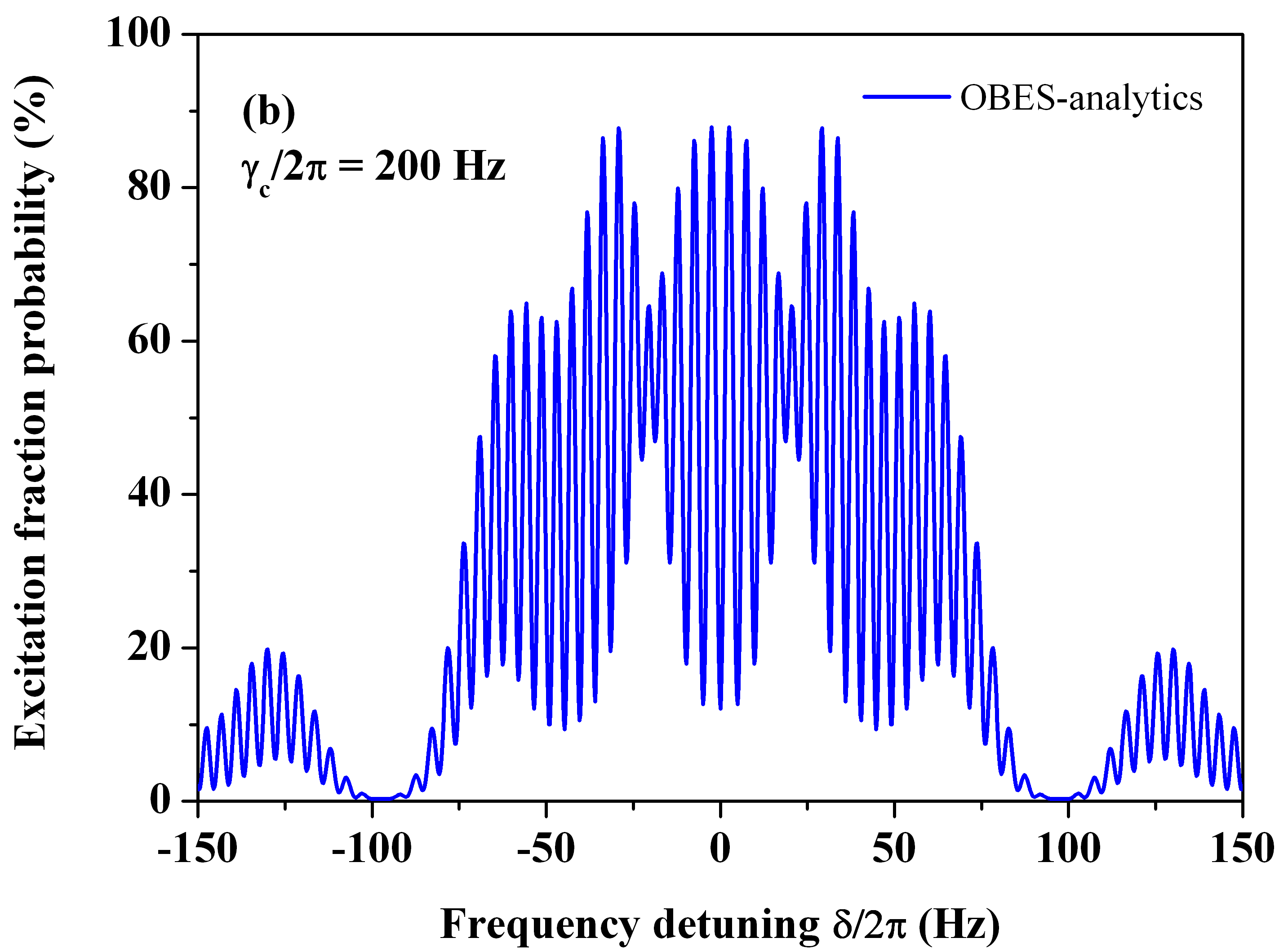}
\includegraphics[width=8cm, height=6cm]{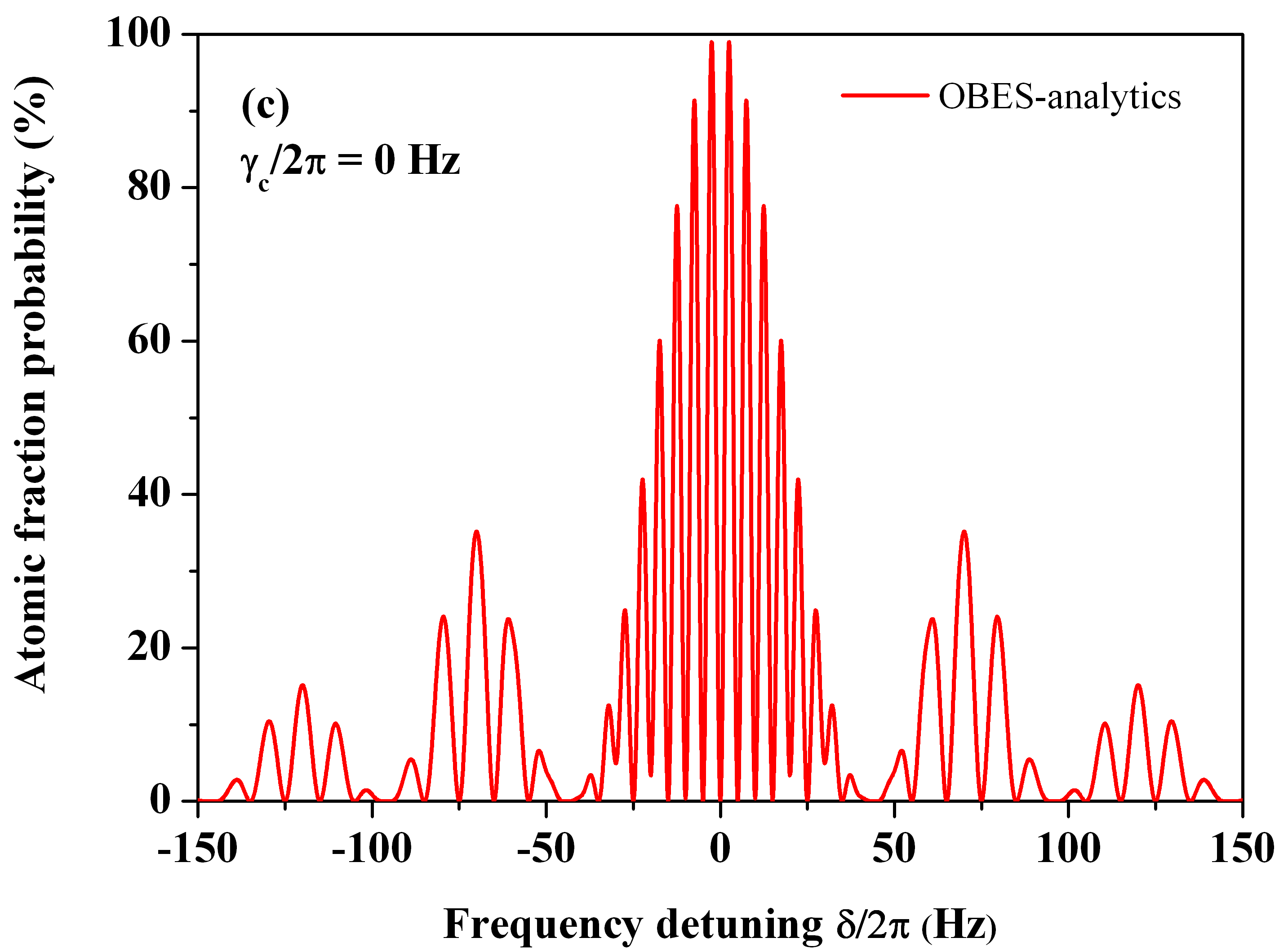}\hspace{1cm}\includegraphics[width=8cm, height=6cm]{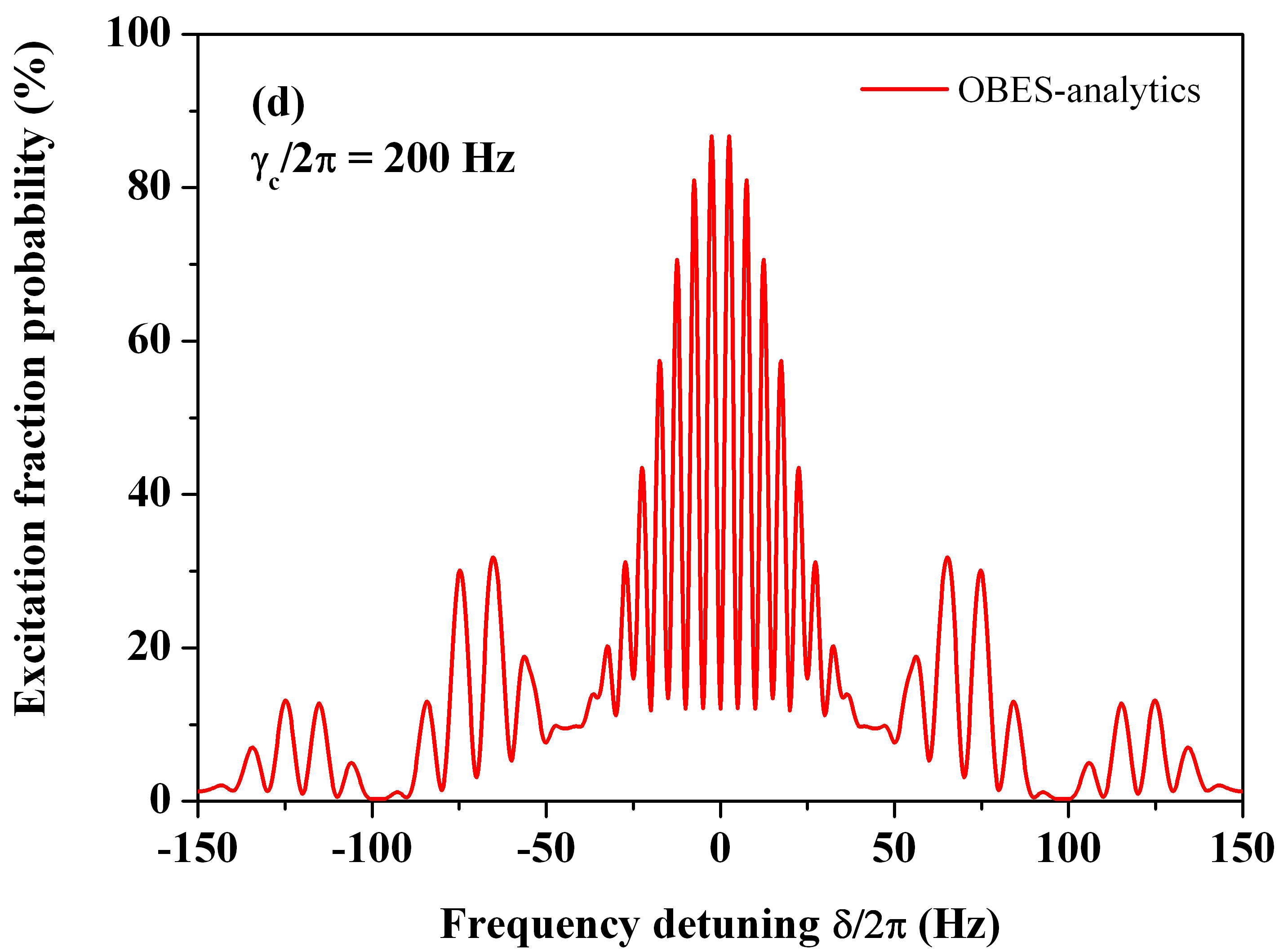}
\caption{Simulations of quantum interferences in the optical clock regime. (a)-(b) Analytical (solid lines) HR3$_{\pi}$ quantum interferences versus the laser probe detuning $\delta/2\pi$. (c)-(d) Analytical (solid lines) DDHR3$_{\pi}$ quantum interferences versus the laser probe detuning $\delta/2\pi$. Laser parameters are $\Omega=\pi/2\tau$ with a pulse duration $\tau=10$~ms and a fixed free evolution time T$=200$~ms. A decoherence term $\gamma_{c}/2\pi=200$~Hz is fixed for graphics (b) and (d). No residual light-shift and no laser probe frequency drift are assumed to be present.}
\label{fig:SM-interferences}
\end{figure*}

In contrast to the probe-frequency offset, the residual light shift is not implemented as an additional phase during the synthetic evolution blocks. For each value of $\Delta_{\rm LS}$, the driven-pulse propagators are recalculated with the corresponding detuning during the pulse segments and recompiled to the IQM native PRX/$R_z$ pulse sequence through the Cayley-Klein mapping of Eq.~\eqref{eq:IQM_gate_app}. The scan therefore probes the light-shift sensitivity of the composite pulse construction at the physical pulse level, whereas $\varepsilon_p$ probes the synthetic free evolution phase response. The probe-frequency offset scan used the same local fitting procedure with 11 evenly spaced values of $\varepsilon_p$ over $\varepsilon_p/2\pi\in[-0.477,0.477]~{\rm MHz}$. and 11 uniformly distributed detuning points per value, with $\varepsilon_p$ implemented through the programmed phase $R_z$ of Eq.~\eqref{eq:IQM_evo_app}. The measured central-fringe shifts versus $\varepsilon_p$ are shown in Fig.~\ref{fig:iqm-pfo}(a).

Finally, quasi-static drift of the probe-frequency offset was emulated by treating $\varepsilon_p$ as a Gaussian-distributed parameter centered at $\varepsilon_p/2\pi = 0.16~{\rm MHz}$ (corresponding to $1$~Mrad/s, chosen to give a measurable contrast loss in HR3$_\pi$) with relative fluctuations $\delta\varepsilon_p/\varepsilon_p = \pm 30\%$. The Gaussian average was approximated using the 21 node Gauss--Hermite quadrature. For each selected value of the effective free evolution time $T$, the contrast was extracted at each quadrature and the final contrast was obtained as the corresponding weighted average. The scan used eight values of $T$ for HR3$_\pi$ and seven values for DDHR3$_\pi$ shown in Fig.~\ref{fig:iqm-pfo}(b). As above, these values of $T$ are effective phase accumulation times, not physical idle times of the qubit. This is why we are able to run the experiment on IQM, going well above the coherence times of the qubit.  Moreover, with all the post processing and the low number of shots, the experiment on IQM took only a couple of minutes.

The residuals from all IQM synthetic-evolution scans are summarized in Table~\ref{tab:iqm_residuals_app}. Probability and contrast residuals are given in percentage points, while central-minimum residuals are given in MHz.

\begin{table}[b]
\centering
\caption{Mean absolute residuals for the IQM synthetic-evolution scans.}
\label{tab:iqm_residuals_app}
\setlength{\tabcolsep}{6pt}
\begin{tabular}{ccccc}
\hline
Var. & Qty. & $N_{\rm sh}$ & HR3$_\pi$ & DDHR3$_\pi$ \\
\hline
$\delta$ & $P_e$ & 1024 & $1.1086~\%$ & $1.1093~\%$ \\
$q$ & $C$ & 128/1024 & $2.222~\%$ & $2.599~\%$ \\
$\Delta_{\rm LS}$ & $\delta_{\rm min}$ & 128 & $0.004~\textrm{MHz}$ & $0.0034~\textrm{MHz}$ \\
$\varepsilon_p$ & $\delta_{\rm min}$ & 128 & $0.0111~\textrm{MHz}$ & $0.0079~\textrm{MHz}$ \\
$\varepsilon_p$ fluct. & $C$ & 128 & $2.13~\%$ & $3.2~\%$ \\
\hline
\end{tabular}
\end{table}

\subsection{Local fitting and minimum extraction}

For the residual-light-shift and probe-frequency offset scans, the central minimum was extracted from a local cosine fit. For each parameter value, the corresponding noiseless reference simulation was used to locate the position of the central minimum, and the experiment sampled a fixed-width detuning window of $\pm 2~{\rm Mrad/s}\,(\equiv\pm 0.318~{\rm MHz})$ around that point. This window width corresponds to a small fraction of the inter-fringe spacing at $T=1~\mu$s, ensuring that exactly one cosine minimum lies in the fitted interval and that no neighboring extrema contaminate the fit. The window is centered around the simulation-predicted minimum, but the cosine fit is performed without any constraint linking its minimum to that position: the experimental minimum is free to lie anywhere within the window. Therefore any displacement of the experimental fringe relative to the simulation prediction is captured by $\delta_{\min}$ as a real shift rather than masked by the window placement. The same procedure was done for the intensity shift fluctuations, but did not yield good results. The shift magnitude is small and due to shot noise error the signal gave bad fits, while the light-shift and probe frequency drift experiments always gave good fits even during testing. Increasing the number of shots to 1024 and having more points per scan did not improve the measured results.

For numerical stability, each local window was fitted in centered coordinates,
\begin{align}
\delta_c=\delta_{\rm MHz}-\delta_{\rm center},\qquad \delta_{\rm center}=\frac{\min(\delta_{\rm sub})+\max(\delta_{\rm sub})}{2},
\end{align}
where $\delta_{\rm sub}$ denotes the detuning points in the local scan window, so $\min(\delta_{\rm sub})$ and $\max(\delta_{\rm sub})$ define the window edges and $\delta_{\rm center}$ is their midpoint. This centering only shifts the data to a local coordinate for the fit; it does not set the reported minimum.

The measured excitation probability was fitted as
\begin{align}
P_e(\delta_c)=A\cos\!\left(2\pi k\delta_c+\phi\right)+B,
\label{eq:local_cos_fit_app}
\end{align}
using constrained nonlinear least squares with $A\geq 0$, $B\geq 0$, and $0\leq\phi\leq 2\pi$. Over the narrow $\pm 0.318$~MHz window, $k$ serves only as a local interpolation parameter. With the convention $A\geq 0$, a local minimum occurs when
\begin{align}
2\pi k\delta_c+\phi=\pi+2\pi n,
\end{align}
so that in centered coordinates
\begin{align}
\delta_{\min}^{c}=\frac{\pi-\phi+2\pi n}{2\pi k},
\label{eq:centered_minimum_app}
\end{align}
where the integer branch $n$ is the unique value placing $\delta_{\min}^c$ inside the fitted window. Transforming back to the original detuning coordinate gives the reported minimum,
\begin{align}
\delta_{\min}=\delta_{\rm center}+\delta_{\min}^{c}.
\label{eq:reported_minimum_app}
\end{align}

\subsection{Finite-shot statistics and interpretation}
\label{sec:iqm_shotnoise}

We summarize agreement through the residuals in Table~\ref{tab:iqm_residuals_app}, rather than adding error bars to every point in the main figures.

As a statistical reference, a probability estimated from $N_{\rm shots}$ binary measurements has the binomial shot-noise scale
\begin{align}
\sigma_P=\sqrt{\frac{P(1-P)}{N_{\rm shots}}}.
\end{align}
The largest value occurs near $P=0.5$, giving $\sigma_P^{\rm max}=1.56\%$ for 1024 shots and $\sigma_P^{\rm max}=4.42\%$ for 128 shots. The corresponding worst-case contrast scales are $1.91\%$ for 1024 shots and $5.41\%$ for 128 shots. In the fluctuating-$\varepsilon_p$ scan, each Gauss--Hermite node has the same $5.41\%$ node-level contrast scale, while the weights of the 21-node Gauss--Hermite average reduce this to approximately $2.37\%$ per final contrast point. For fitted minima, the conversion of probability shot noise into an uncertainty in $\delta_{\min}$ depends on the local fringe curvature and nonlinear fit, so the residuals are reported directly in MHz. The slightly larger DDHR3$_\pi$ residuals in the pulse-area and $\varepsilon_p$-fluctuation contrast scans (Table~\ref{tab:iqm_residuals_app}) sit modestly above the worst-case $P=0.5$ binomial bounds.

\section{Analytical time-dependent SO(3) matrix elements including decoherence}

\indent We provide analytical matrix solutions of laser pulses used to design sequences of phase-shifted refocusing pulses when decoherence is present.
Generalized hyper-Ramsey and dynamically-decoupled hyper-Ramsey protocols are described within the density matrix formalism.
The optical Bloch equations presented in the main text describe the laser field interaction with a two-state quantum system. The general solution $\textup{M}(\theta_{l})$ is derived in a matrix form written as~\cite{Schoemaker:1978-SM,Zanon-Willette:2017-SM}:
\begin{equation}
\begin{split}
&\textup{M}(\theta^{\pm}_{\varphi_{l}})\cdot\left(\begin{array}{c}
                              C_{gg}(0) \\
                              C_{ee}(0)
                            \end{array}\right)\rightarrow\textup{R}(\theta^{\pm}_{\varphi_{l}})\textup{R}(0)\\
&\textup{R}(\theta^{\pm}_{\varphi_{l}})\textup{R}(0)=\left(
  \begin{array}{ccc}
  \textup{R}_{11}(\theta^{\pm}_{\varphi_{l}}) & \textup{R}_{12}(\theta^{\pm}_{\varphi_{l}}) & \textup{R}_{13}(\theta^{\pm}_{\varphi_{l}}) \\
 \textup{R}_{21}(\theta^{\pm}_{\varphi_{l}}) & \textup{R}_{22}(\theta^{\pm}_{\varphi_{l}}) & \textup{R}_{23}(\theta^{\pm}_{\varphi_{l}}) \\
 \textup{R}_{31}(\theta^{\pm}_{\varphi_{l}}) & \textup{R}_{32}(\theta^{\pm}_{\varphi_{l}}) & \textup{R}_{33}(\theta^{\pm}_{\varphi_{l}}) \\
  \end{array}
 \right)\cdot\left(
 \begin{array}{c}
 \textup{U}(0) \\
 \textup{V}(0)  \\
 \textup{W}(0)  \\
 \end{array}
 \right)
,\\
\end{split}
\label{Jaynes-solution}
\end{equation}
where the clock frequency detuning is defined by $\delta_{l}=\pm(\delta-\Delta_{l})+\varepsilon_{P}$. The generalized pulse area is $\theta_{l}=\omega_{l}\tau_{l}$.
These square matrix elements $\textup{R}_{\textup{mn}}(\theta_{l})$ following~\cite{Zanon-Willette:2017-SM} are given by:
\begin{equation}
\begin{split}
\textup{R}_{11}(\theta_{l})=e^{-\gamma_{c}\tau_{l}}&\left(a_{0}-a_{2}[\delta_{l}^{2}+\Omega_{l}^{2}\sin^{2}\varphi_{l}]\right),\\
\textup{R}_{12}(\theta_{l})=e^{-\gamma_{c}\tau_{l}}&\left(a_{1}\delta_{l}+a_{2}\Omega_{l}^{2}\sin\varphi_{l}\cos\varphi_{l}\right),\\
\textup{R}_{13}(\theta_{l})=e^{-\gamma_{c}\tau_{l}}&\left(a_{2}[\delta_{l}\Omega_{l}\cos\varphi_{l}-\Delta\gamma\Omega_{l}\sin\varphi_{l}]\right.\\
&\left.-a_{1}\Omega_{l}\sin\varphi_{l}\right),\\
\textup{R}_{21}(\theta_{l})=e^{-\gamma_{c}\tau_{l}}&\left(-a_{1}\delta_{l}+a_{2}\Omega_{l}^{2}\sin\varphi_{l}\cos\varphi_{l}\right),\\
\textup{R}_{22}(\theta_{l})=e^{-\gamma_{c}\tau_{l}}&\left(a_{0}-a_{2}[\delta_{l}^{2}+\Omega_{l}^{2}\cos^{2}\varphi_{l}]\right),\\
\textup{R}_{23}(\theta_{l})=e^{-\gamma_{c}\tau_{l}}&\left(a_{2}[\delta_{l}\Omega_{l}\sin\varphi_{l}+\Delta\gamma\Omega_{l}\cos\varphi_{l}]\right.\\
&\left.+a_{1}\Omega_{l}\cos\varphi_{l}\right),\\
\textup{R}_{31}(\theta_{l})=e^{-\gamma_{c}\tau_{l}}&\left(a_{2}[\delta_{l}\Omega_{l}\cos\varphi_{l}+\Delta\gamma\Omega_{l}\sin\varphi_{l}]\right.\\
&\left.+a_{1}\Omega_{l}\sin\varphi_{l}\right),\\
\textup{R}_{32}(\theta_{l})=e^{-\gamma_{c}\tau_{l}}&\left(a_{2}[\delta_{l}\Omega_{l}\sin\varphi_{l}-\Delta\gamma\Omega_{l}\cos\varphi_{l}]\right.\\
&\left.-a_{1}\Omega_{l}\cos\varphi_{l}\right),\\
\textup{R}_{33}(\theta_{l})=e^{-\gamma_{c}\tau_{l}}&\left(a_{0}+a_{1}\Delta\gamma-a_{2}[\Omega_{l}^{2}-\Delta\gamma^{2}]\right),
\end{split}
\label{Bloch-matrix-components}
\end{equation}
where $\Delta\gamma=\gamma_{c}$.
Auxiliary time-dependent functions $\textup{a}_{0}\equiv\textup{a}_{0}(\theta_{l}),\textup{a}_{1}\equiv\textup{a}_{1}(\theta_{l}),\textup{a}_{2}\equiv\textup{a}_{2}(\theta_{l})$ are given in~\cite{Zanon-Willette:2017-SM}:
\begin{equation}
\begin{split}
a_{0}(\theta_{l})=&[(\textup{SD}_{3}-\textup{TD}_{2})\sin\theta_{l}+(\textup{SD}_{2}+\textup{TD}_{3})\cos\theta_{l}]e^{\rho_{l}\tau_{l}}\\
&+(\textup{D}_{0}\eta_{l}+\textup{g}_{l}^{2})\textup{R}e^{\eta_{l}\tau_{l}},\\
a_{1}(\theta_{l})=&[(\textup{SD}_{1}-\textup{T}\omega_{l})\sin\theta_{l}+(\textup{S}\omega_{l}+\textup{TD}_{1})\cos\theta_{l}]e^{\rho_{l}\tau_{l}}\\
&+\textup{D}_{0}\textup{R}e^{\eta_{l}\tau_{l}},\\
a_{2}(\theta_{l})=&[\textup{S}\sin\theta_{l}+\textup{T}\cos\theta_{l}]e^{\rho_{l}\tau_{l}}+\textup{R}e^{\eta_{l}\tau_{l}},
\end{split}
\label{auxiliary-functions}
\end{equation}
We introduce the following notation:
\begin{equation}
\begin{split}
g_{l}^{2}=&\Omega_{l}^{2}+\delta_{l}^{2},\\
\textup{D}_{0}=&\eta_{l}-\Delta\gamma,\\
\textup{D}_{1}=&\rho_{l}-\Delta\gamma,\\
\textup{D}_{2}=&\omega_{l}(2\rho_{l}-\Delta\gamma),\\
\textup{D}_{3}=&(\rho_{l}^{2}-\omega_{l}^{2}-\rho_{l}\Delta\gamma+g_{l}^{2}),\\
\end{split}
\end{equation}
and
\begin{equation}
\begin{split}
\textup{R}=&\frac{1}{(\rho_{l}-\eta_{l})^{2}+\omega_{l}^{2}},\\
\textup{S}=&\frac{(\rho_{l}-\eta_{l})}{\omega_{l}((\rho_{l}-\eta_{l})^{2}+\omega_{l}^{2})},\\
\textup{T}=&\frac{-1}{(\rho_{l}-\eta_{l})^{2}+\omega_{l}^{2}}.
\end{split}
\label{integration-constants}
\end{equation}
The three roots of the matrix (one real root $\eta_{l}$ and two complex ones $\rho_{l}\pm i\omega_{l}$) are by Cardan's cubic solutions leading to damping terms $\eta_{l},\rho_{l}$
and a generalized angular frequency $\omega_{l}$ written as:
\begin{equation}
\begin{split}
\eta_{l}=&\frac{1}{3}(\Delta\gamma-\textup{C}-\frac{\Delta_{0}}{\textup{C}}),\\
\rho_{l}=&\frac{1}{3}(\Delta\gamma+\frac{\textup{C}}{2}+\frac{\Delta_{0}}{2\textup{C}}),\\
\omega_{l}=&\frac{\sqrt{3}}{6}(-\textup{C}+\frac{\Delta_{0}}{\textup{C}}),\\
\Delta_{0}=&\Delta\gamma^{2}-3g_{l}^{2},\\
\Delta_{1}=&-2\Delta\gamma^{3}+9g_{l}^{2}\Delta\gamma - 27\delta_{l}^{2}\Delta\gamma,\\
\textup{C}=&\sqrt[3]{\frac{\Delta_{1}+\sqrt{\Delta_{1}^{2}-4\Delta_{0}^{3}}}{2}}.
\end{split}
\label{cardan-roots}
\end{equation}
The free evolution matrix $\textup{R}(\pm)$ without laser field reduces to:
\begin{equation}
\begin{split}
\textup{R}(\pm)&=\prod\limits_{m=2\mathbb{N}^{*}}\sqrt[m]{\textup{R}(\pm)}\\
&=e^{-\gamma_{c}\textup{T}}\left(
\begin{array}{ccc}
\cos\left(\pm\delta+\varepsilon_{P})\textup{T}\right) & \sin\left(\pm\delta+\varepsilon_{P})\textup{T}\right) & 0 \\
-\sin\left(\pm\delta+\varepsilon_{P})\textup{T}\right) & \cos\left(\pm\delta+\varepsilon_{P})\textup{T}\right) & 0\\
0 & 0 & 1
\end{array}\right).
\end{split}
\label{free-rotation-matrix-DDHR}
\end{equation}

\end{document}